\documentclass[11pt]{article}
\pdfoutput=1
\usepackage{jheppub}
\usepackage{graphicx}
\usepackage{amssymb, amsmath, amssymb}
\usepackage{slashed}
\usepackage{cancel}
\usepackage{hyperref}
\usepackage{caption}
\usepackage{xcolor}
\usepackage{dsfont}
\usepackage{verbatim}
\usepackage{subfig}
\usepackage{float}
\usepackage{braket}
\usepackage{longtable}
\usepackage{array}
\usepackage{tikz-feynman} 
\usetikzlibrary{external}
\tikzfeynmanset{compat=1.1.0}
\tikzset{double line/.style={double distance=10pt,draw}}
\usetikzlibrary{decorations.markings}
\usepackage{rotating}
\usepackage{lscape}
\usetikzlibrary{chains}

\newcommand\beq{\begin{equation}}
\newcommand\eeq{\end{equation}}
\newcommand\be{\begin{equation}}
\newcommand\ee{\end{equation}}

\newcommand{\expt}[1]{\left< #1 \right>}
\newcommand{\Tr}[1]{\text{Tr}\left( #1 \right)}
\newcommand\ra{\rightarrow}

\title{The late time ramp from chord diagrams in the double-scaled SYK model}

\preprint{UT-WI-10-2025}

\author{Amir Raz, }
\author{Merna Youssef}

\affiliation{Theory Group, Weinberg Institute, Department of Physics, University of Texas\\
2515 Speedway, Austin, TX 78712, USA.}

\emailAdd{araz@utexas.edu}
\emailAdd{myoussef@utexas.edu}

\abstract{We compute the ramp of the spectral form factor analytically from chord diagrams in double scaled SYK. We map the double-trace correlator to a sum of single trace two-point functions over a basis of operators. We then reproduce the local eigenvalue correlations in random matrix theory from the chord diagrams perspective, which is the $q= 0$ limit of double scaled SYK, and identify the relevant operators that give rise to the late-time ramp. We then extend the computation to finite $q$, resulting in the late time contribution to the spectral form factor. We verify that the late time asymptotics of the finite $q$ computation gives rise to the expected late time ramp. Our computation also provides the corresponding trumpet partition function and gluing factor for chords, which form the basis of a chord analog to topological recursion.}

\begin{document}
\maketitle

\section{Introduction}
In recent years, the SYK model has received a lot of attention as a model that could potentially describe the dynamics of black holes \cite{Kitaev_talk,MaldecenaStanford,Kitaev_2018,Kitaev_2019,Rosenhaus_2019}. It is a quantum mechanical model of $N$ Majorana fermions with all-to-all interaction and disordered random couplings. This model has stark similarities to a theory of gravity; in the large coupling limit and the large $N$ limit it has a conformal regime in the IR with an emergent reparametrization symmetry SL(2,R) the same symmetry breaking pattern in the cutoff versions of Euclidean $AdS_2$ spacetime\cite{NADSbreaking,MaldecenaStanford, Kitaev_talk}. Moreover, the model has a maximal Lyapunov exponent \cite{Jensen_2016} which is also exhibited by nearly extremal black holes \cite{chaos,MaldecenaStanford}. These similarities has opened the door to understand black hole physics from the point of view of quantum chaos. 

There are many different probes of quantum chaos in the physics and quantum information literature. For example, the decay of the out of time ordered correlators \cite{chaos,Stanford_2022}, the growth of operator complexity \cite{Roberts_2018}, and the local eigenvalue correlations \cite{RMTconcepts,BGSconj,You_2017,Garc_a_Garc_a_2016,Cotler:2016fpe} are all used to identify if a system is chaotic, and to quantify chaos. In this article, we will focus on the last one, namely that the distribution of nearest eigenvalue spacings of a chaotic Hamiltonian follow a Wigner-Dyson distribution. In practice, it is not easy to probe the eigenvalue pair correlator directly, and instead it is simpler to compute double trace correlators like the spectral form factor.

Given an ensemble of Hamiltonians, one can compute many ensemble averaged observables. One particular observable of interest is the average eigenvalue distribution
\begin{equation}
    \rho(E) = \expt{\frac{1}{L}\sum_{i=1}^L\delta(E - E_i)} = \frac{1}{\Tr{1}} \expt{\Tr{\delta(E - H)}} ,
\end{equation}
where the brackets $\expt{\text{---}}$ denote an expectation value over the ensemble, and $L=\Tr{1}$ is the size of the Hilbert space the Hamiltonian acts on. From the average eigenvalue distribution one can compute many other single-trace observables, as
\begin{equation} \label{eq:sngltr}
    \frac{1}{\Tr{1}}\expt{\Tr{f(H)}} = \int dE ~\rho(E) f(E),
\end{equation}
for all (sufficiently well behaved) functions $f(x)$. Instead of computing the average eigenvalue distribution, one can instead compute the partition function
\begin{equation}
    Z(\beta) = \frac{1}{\Tr{1}}\expt{\Tr{e^{-\beta H}}} = \int dE ~\rho(E) e^{-\beta E}.
\end{equation}
The average density of states, $\rho(E)$, and the ensemble averaged partition function $Z(\beta)$, are related by a Laplace transform, and so both contain the same information about the ensemble.

Many ensembles of interest, like the ensembles in classical random matrix theory or the SYK model, are self averaging. This means that a typical member in the ensemble resembles the ensemble average. For example, if an ensemble is self averaging then the partition function of a Hamiltonian picked at random will only deviate slightly from the ensemble averaged partition functions, with these deviations dying out in the $L \ra \infty$ limit. This allows one to draw conclusions about a typical Hamitonian, even if no ensemble averaging is preformed.

A similar story holds for correlations between eigenvalues. One can compute the average pair density
\begin{equation}
    \rho(E,E') = \frac{1}{\Tr{1}^2} \expt{\Tr{\delta(E - H)} \Tr{\delta(E' - H)}} .
\end{equation}
For energies separated by a finite distance we expect this density to factorize if the ensemble is self averaging, namely $\rho(E,E') \approx \rho(E) \rho(E')$ for $E-E' = O(1)$. However, when the energies are extremely close and of the order of the level spacing or $L^{-1}$, we expect to see the universal eigenvalue repulsion of random matrix theory.

Just as the average density can be used to compute single trace observables, the average pair-density correlator can be used to compute double-trace observables. Of particular interest to us are the double-trace partition function
\begin{equation}
    G(\beta_1,\beta_2) =  \frac{1}{\Tr{1}^2} \expt{\Tr{e^{-\beta_1 H}} \Tr{e^{-\beta_2 H}}} =  \int dE_1~dE_2~\rho(E_1,E_2) e^{-\beta_1 E_1 - \beta_2 E_2},
\end{equation}
and the spectral form factor (SFF)
\begin{equation} \label{eq:SFF}
    g(\beta,t) =  G(\beta + it,\beta - it)
    = \int dE_1~dE_2~\rho(E_1,E_2) e^{-\beta( E_1 +E_2) - i t(E_1- E_2)} .
\end{equation}
Just as the partition function contains the same information as the average spectral density, the SFF contains the same information as the average pair-density correlator. In particular at very late times $t = O(L)$, the integral in \eqref{eq:SFF} localizes $\rho(E_1,E_2)$ near $E_1-E_2 = O(L^{-1})$, and so the very late time behavior of the SFF probes the universal eigenvalue repulsion of RMT. This results in a universal ramp in the SFF. At even later times the SFF plateaus due to the underlining discreteness of the spectrum \cite{Br_zin_1997,Cotler:2016fpe,Liu_2018}.

The SFF is also a useful tool in holography. In the boundary theory the SFF can be expressed as a sum over the energy eigenstates, as in \eqref{eq:SFF}. This correlator oscillates erratically at late times due to the energy phase factors, and only terms with almost the same energies will dominate giving a non zero value. But what is the analogous behavior in the gravity side? In JT gravity, double trace correlators correspond to a path integral over topologies with two asymptotic boundaries \cite{saad2018,saad2019late,Saad:2019lba,Witten:2020wvy,Mertens_2023}. The leading contribution to this correlator comes from a disconnected topology, which decays at late times. The late time RMT ramp comes from adding contributions from a connected double-trumpet topology (i.e. a spacetime wormhole) \cite{Saad:2019lba}. In fact, JT gravity is exactly dual to a particular double-scaled matrix model \cite{Saad:2019lba,stanford2019jt}, and the universality of the ramp comes from the universal form of the double-resolvent in the topological expansion of random matrix theory.

While the Schwarzian limit of the SYK model is dual to JT gravity, the full SYK model also contains a tower of massive modes \cite{Rosenhaus_2019,Polchinski_2016,Gross_2017,MaldecenaStanford}. Nevertheless, it still has the universal ramp in the SFF coming from the universal RMT correlations between local eigenvalues \cite{Cotler:2016fpe,verbaarschot2019,Garc_a_Garc_a_2016,Kanazawa_2017,Altland_2018}. Many attempts have been made to find the universal ramp in disconnected saddles \cite{saad2018,Aref_eva_2019,Wang_2019,Khramtsov_2021}, or through other means \cite{Altland_2018,Behrends2020,Behrends_2020,verbaarschot2019}, though none provide a clean gravitational interpretation like in JT gravity. This is further complicated by the fact that the SYK model also contains global correlations between the eigenvalues, which are much more dominant at early times \cite{Altland_2018,verbaarschot2019,Berkooz2021,Cotler:2016fpe,C_ceres_2022}. These global fluctuations are suppressed by a power law in $N \sim \log(L)$, where the dimension of the SYK Hilbert space is $L = 2^{N/2}$, while the universal topological contribution is exponentially suppressed, scaling as $L^{-2} \sim 2^{-N}$. Understanding the interplay between these distinct contributions and the gravitation dual, remains elusive.

One solvable limit of the SYK model is the double-scaled limit \cite{Micha2018,Berkooz_2019,berkooz2020complex,Berkooz_2020,Lin_2022,Lin_2023}.\footnote{For a recent review of DSSYK see \cite{berkooz2024cordial}} In this limit the number of fermions in the interaction term grows as $\sqrt{N}$ as we take $N\ra \infty$. Interestingly, this limit smoothly interpolates between pure random matrix theory when the number of fermions in the interaction term are of order $N$, and the standard SYK model where the interaction length is finite as $N\ra\infty$ \cite{Erdos14,feng2018spectrum,Micha2018,Berkooz_2019}. In this limit, single trace operators and correlation functions, ala \eqref{eq:sngltr}, can be computed using the machinery of chord diagrams, which we will review in section \ref{sec:CD}. While this technique is well established for single trace observables, it has not really been developed from first principles for the universal piece of the double-trace correlator. \cite{Berkooz2021} used chords to compute the leading global fluctuations, while \cite{okuyama2023endworldbranedouble} stated what the universal piece in the dual gravitational Hilbert space should be, rather than computing it from chord diagrams. Other dual descriptions of the double scaled SYK model (DSSYK) correctly match single trace observables, but either do not consider matching double-trace correlators, or assume the universal correlations in double-trace observables, like in \cite{aguilar2025}. One prominent supposed gravitational dual of DSSYK is sine-dilaton gravity \cite{Blommaert_2025,bossi2025sine,blommaert2025}, which has been shown to be dual to a single cut random matrix theory with a $q$-Gaussian spectrum \cite{blommaert2025}.\footnote{This is not surprising as sine dilaton gravity is a potential deformation of JT gravity, and potential deformations of JT gravity have been shown to be dual to single cut matrix models with a deformed spectrum in \cite{Witten:2020wvy}.} Thus the duality between DSSYK and sine dilaton gravity is exact for single trace observables, but misses the global fluctuations, or sparseness, present in spectral fluctuations of the DSSYK model.

The goal of this work is to construct the universal RMT contribution to the SFF of the DSSYK model directly from chord diagrams. Our approach is to start from the RMT limit of DSSYK, where the universal contribution and topological recursion are well established, and then extend our results to the full DSSYK.

We begin with a review of chord diagrams and the DSSYK in section \ref{sec:CD}. We then review double-trace correlators in the SYK model, and the different time scales of the SFF in section \ref{sec:DTinSYK}. In section \ref{chords in rmt} we compute the universal eigenvalue correlator in RMT using non-intersecting chord diagrams on the cylinder. We then extend this computation by allowing some intersecting chords in section \ref{fromrmttofiniteq}, resulting in a computation of the late time ramp in DSSYK.

\section{Review of double scaled SYK and Chord Diagrams} \label{sec:CD}

In this section we review the double scaled limit of the SYK model, and the chord diagram approach to evaluating observables like the partition function, following \cite{Berkooz_2019,Micha2018}. Readers familiar with this methodology may wish to skip to section \ref{sec:DTinSYK}.

The SYK model is given by $N$ Majorana fermions $\psi_i$, $i\in \{1, \ldots, N\}$, with the canonical commutation relations $\{\psi_i ,\psi_j\} = 2 \delta_{i,j}$. It is convenient to define for each ordered index set $I= \{i_1,i_2,\ldots,i_{I_p}\} \subset \mathbb{Z}_N$, the multi-fermion Hermitian operator:
\begin{equation}
    \Psi_I \equiv i^{p_I(p_I-1)/2} \psi_{i_1} \psi_{i_1} \cdots \psi_{i_{p_I}}.
\end{equation}
Note that with this normalization $\Psi_I^2 = \mathbf{1}$, the identity operator.

Then the Hamiltonian for the SYK model of length $p$ is 
\begin{equation}
    H = \sum_{I \subset \mathbb{Z}_N, |I| = p} J_I \Psi_I,
\end{equation}
where $J_I$ are identical random Gaussian coupling with mean zero and variance
$\expt{J_I J_{I'}} = \binom{N}{p}^{-1} \delta_{I,I'}$. Under this convention we have that $\Tr{H^2} = \Tr{\mathbf{1}}$. We will assume throughout that $N$ and $p$ are even unless mentioned otherwise.

In the double scaling limit we take $N,p \ra \infty $, while keeping the ratio $\lambda \equiv \frac{2p^2}{N}$ fixed. The density of states of this model converges in probability to a Q-Gaussian distribution with parameter $q = e^{-\lambda}$, which we will compute using chord diagrams.

\begin{figure}
    \centering
    \includegraphics[width=.7\textwidth]{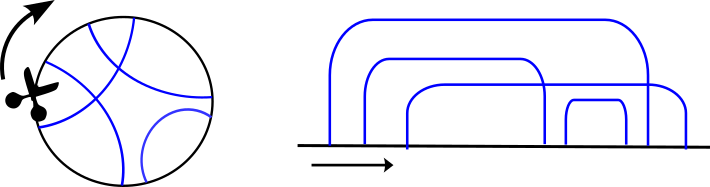}
    \caption{a Chord diagram on the left with 8 nodes representing 8 copies of the Hamiltonian. Each node has $p$-interacting fermions with associated random coupling. On the right, we open this chord diagram into a straight line.}
    \label{fig:cutting chord diagram}
\end{figure}

To compute the density of states, we start by analyzing the partition function of the model: $Z = \langle \Tr{e^{-\beta H}}\rangle_J / \Tr{\mathbf{1}}$ which can be written as taylor series in H such that:
\begin{equation}
Z = \sum_{k=0}^{\infty} \frac{(-\beta)^k}{k!} \frac{\langle \Tr H^k \rangle}{\Tr{\mathbf{1}}} = \sum_{k=0}^{\infty} \frac{(-\beta)^k}{k!} m_k  
\end{equation}
In the chord diagram prescription, we model the trace by a circle with nodes. For instance $m_k = \langle \Tr H^k \rangle / \Tr{\mathbf{1}}$ will have a circle with $k$ nodes, each represents a copy of the Hamiltonian with one single random coupling constant $J$. Each node is assigned an index $j=1, j=2, j=3...j=k$, and has string of length $p$, i.e $\Psi_{I_1} = \{ \psi_{i_1} \psi_{i2}...\psi_{i_p} \}$ where $I_j$ represents the set of $p$-randomly picked interacting fermions with a random variable associated to this string $J_{I_1}$. See Figure \ref{fig:cutting chord diagram}. 

\begin{figure}
    \centering
    \includegraphics[width=.7\textwidth]{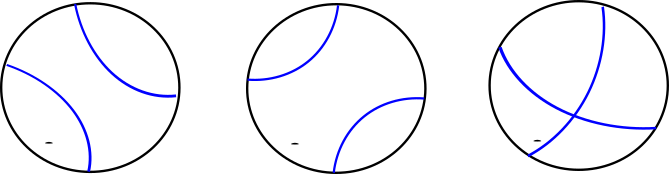} 
    \caption{The Chord diagram interpretation of $m_4$ amounts to 3 configurations: two of them with no intersections and the last one with one intersection $=2+q$, the same value we are getting from wick contraction.}
    \label{fig:m4}
\end{figure}

The moment takes the form:
\begin{equation}
    m_k   = \frac{\langle Tr H^k \rangle}{\Tr{\mathbf{1}}} =  i^{\frac{kp}{2}} \sum_{\{I_1 ..I_k\}} \langle J_{I_1}...J_{I_k}\rangle \Tr{\Psi_{I_1}...\Psi_{I_k}}
    \label{moment}
\end{equation} 
where the moments of J (the ensemble average over the J's) will take the values of ${N \choose p}^{-k/2}$ so that will give 
\begin{equation}
    m_k  = i^{\frac{kp}{2}} \sum_{\{I_1 ..I_{k}\}} {N \choose p} ^{-k/2}  \Tr{\Psi_{I_1}...\Psi_{I_k}}
    \label{ensavgoverj}
\end{equation}
since the $J's$ are Gaussian variables, Wick's theorem identifies pairs of indices of $\Psi$. This means that each $I_j$ will appear exactly twice so the number of distinct elements of $\{ I_j\}$ is $k/2$. Thus, the wick contraction of the moments of $J$ in equation \ref{moment} identifies which pairs of $\Psi$'s have the same index. For example, let's compute $m_4 = \langle \Tr H^4 \rangle$ disregarding the constants for the time being, we have:
\begin{equation}
\begin{aligned}\\&
    m_4 = i^{\frac{4p}{2}} \sum_{\{I_1 ..I_4\}} \langle J_{I_1} J_{I_2} J_{I_3} J_{I_4} \rangle \Tr {\Psi_{I_1}\Psi_{I_2}\Psi_{I_3}\Psi_{I_4}}\\& = \# \sum_{\{I_1 ..I_4\}}(\delta_{I_1 I_2}\delta_{I_3 I_4} + \delta_{I_1 I_4}\delta_{I_2 I_3}+\delta_{I_1 I_3}\delta_{I_2 I_4})\Tr {\Psi_{I_1}\Psi_{I_2}\Psi_{I_3}\Psi_{I_4}}\\&
    =\# \sum_{\{I_1 ..I_4\}} (\Tr {\Psi_{I_1}\Psi_{I_1}\Psi_{I_3}\Psi_{I_3}}+\Tr {\Psi_{I_1}\Psi_{I_2}\Psi_{I_2}\Psi_{I_1}}+\Tr {\Psi_{I_1}\Psi_{I_2}\Psi_{I_1}\Psi_{I_2}})
\end{aligned}
\end{equation}
These three contribution are presented pictorially in figure \ref{fig:m4}. To evaluate these traces we need to permute the $\Psi$'s to get the ones with similar indices adjacent to each other and annihilate them. Commuting $\Psi_{I_1} \& \Psi_{I_2}$ will give a factor of $(-1)^{I_1 \cap I_2}$ the intersection of two sets is Poisson distributed in the double scaling limit \cite{Erdos14}:
the probability of having an m fermions shared by the two sets is:
\begin{equation}
\begin{aligned}\\&
    P({I_1 \cap I_2} = m) = {p \choose m}{N-p \choose p-m}/{N \choose p} = \frac{(p^2/N)^m}{m!}e^{-p^2/N}(-1)^m \\&
    \implies \sum_{\{I_1 ..I_k\}} (-1)^{{I_1 \cap I_2}}=\sum_{m=0}^
    \infty \frac{(p^2/N)^m}{m!}e^{-p^2/N}(-1)^m = e^{-\lambda}\equiv q
\end{aligned}
\end{equation}
where the sum over $m$ is going to infinity in the double scaling limit. See Figure 2 for interpreting this into chord diagrams.

Now equation \ref{ensavgoverj} with the factor of $ {N\choose p}^{-k/2}$ turns the appearance of each term in the sum into a probability of such an event, hence eq. \ref{moment} could be written as:
\begin{equation}
    m_k = \sum_{\substack{\text{chord diagrams}\\ \text{of $k/2$ chords}}} q^{\#\text{intersections}}
    \label{sumovercd}
\end{equation}
In other words, the ensemble average of the moment $m_k$ reduces to a counting of chord diagrams. Eq.\ref{sumovercd} is the chord partition function. 
\subsection{From the Hamiltonian to the Transfer matrix}
For higher powers of the Hamiltonian, drawing these diagrams won't be practical so we need to translate the summation over chord diagrams to a recursion relation that generates every possible diagram for a given $k$. To visualize it we open the chord diagram as in Figure 1. the detailed explanation can be viewed in \cite{Micha2018,Berkooz_2019}. We briefly explain it here. Pick a point on the chord diagram, cut it open and move clockwise. The first node you hop over marks the first "step". the number of $H$ factors we hop over will be denoted by $i$. such that at the zeroth step, every chord is already paired with another. At every step, a chord can close or a new chord can emanate from it such that the chords that emanate have to be lower than those on its left (to make sure chords don't intersect unnecessarily when closing them). Denote by $v_l^{(i)}$ the open or partial chord partition function: the sum over chord configurations with l open chords at step i weighted by intersections of closing chords at step $i$.
\begin{equation}
    v_l^{(i)} = \sum_{\pi \in \Pi(i,l)} q ^{k_p(\pi)}
\end{equation}
where $k_p(\pi)$ is the number of intersections happened to the left of step $i$. Now one can write a recursion relation for $v_l^{i+1}$:
\begin{equation}
    v_l^{i+1} = v_{l-1}^{(i)}+(1+q+...+q^l)v_{l+1}^{(i)}
    \label{recursionrelation}
\end{equation}
In other words: the open chord partition function with l open chords at step $i+1$ means that you either had l+1 open chord at step $i$ and one of them closed at step $i+1$ (giving a contribution of $q^{\# \text{ intersections on its way down with the other chords}}$) or you had $l-1$ open chords at step $i$ and another has opened up. This recursion can be written in terms of an $(L+1)\times(L+1)$ transfer matrix with indices running from 0 to $L$:
\begin{equation}
v^{(i+1)} = T_Lv^{(i)} \,\, \text{where} \,\,\,
    \left[T_L\right]_{l_1}^{l_2} = \delta_{l_1-1}^{l_2}+\eta_{l_1}\delta_{l_1+1}^{l_2}
    \label{recursionrelation2}
\end{equation}
where $l_1$ is the row index, $l_2$ is the column index and $\eta_l = 1+q+...+q^l = \frac{1-q^{l+1}}{1-q}$. This matrix will take the form:
\begin{equation}
T_{(L)}=
{\begin{bmatrix}
0 &  \frac{1-q}{1-q}& 0 & 0 & 0&\cdots\\
1 & 0 & \frac{1-q^2}{1-q} & 0  &0 &\cdots\\
0 & 1 &0& \frac{1-q^3}{1-q}  &0&\cdots\\
\vdots & \ddots &\ddots&&\ddots &\cdots \\
\end{bmatrix}}_{(L+1)\times (L+1)}
\label{transfermatrix}
\end{equation}
the computation of the chord partition function amounts to projecting this matrix on the initial condition vector L times.
\begin{equation}
    \ket{0}_L = v^{(0)} = (1,0,...,0)^T
\end{equation}
with entries $L+1$. Such that 
\begin{equation}
    m_L = \langle 0|T^L|0 \rangle
    \label{momenttransfer}
\end{equation} 
To illustrate: if we calculate $m_4$ by projecting the transfer matrix 4 times on the initial vector, we obtain $2+q$ as expected from the counting argument outlaid before. In other words, it is a matter of calculating the spectrum of T. denote the eigen values of $T$ by $\alpha$:
\begin{equation}
    m_L = \int_{\text{Spec(T)}} d\alpha \,\, \alpha^L \,\, \rho(\alpha) \,\, |\psi(\alpha)|^2 
\end{equation}
where $\rho(\alpha)$ is the eigen value density of $T$ and $\psi _0(\alpha)  = \braket{0|\alpha}$ is the overlap of $\bra{0}$and $\ket{\alpha}$ eigen vectors. Comparing the original moment in \ref{moment} and \ref{momenttransfer}, we identify $E = \alpha$ where $v(E|q) = \rho(E) \,\, |\psi_0(E)|^2 $ constitute the asymptotic distribution of the energies (probability density of eigen values in the large L limit). \ref{transfermatrix} is a Toeplitz tridiagonal matrix (with constant elements one diagonal above and below the main diagonal). The eigen values for such matrices are well known 
\begin{equation}
  \frac{-2}{\sqrt{1-q}}\cos \frac{s \pi}{L+1} = \frac{-2}{\sqrt{1-q}} \,\, \cos \theta, \,\, \text{where $s$ runs over $1,...L$}  
\end{equation}
The support of the eigen values is $(\frac{-2}{\sqrt{1-q}},\frac{2}{\sqrt{1-q}})$. In the infinite $L$ limit, we have a continuous spectrum where $\theta$ covers the interval $\left[0,\pi\right]$. Next step is to find the corresponding eigen vectors. By parametrizing the eigen values of T as $E(\mu) \equiv \frac{2\mu}{\sqrt{1-q}}$ with $v^{(\mu)}$ the corresponding eigen vector. The recursion relation eq \eqref{recursionrelation} and eq \eqref{recursionrelation2} can be written as 
\begin{equation}
T v^\mu = \frac{2\mu}{\sqrt{1-q}}v^\mu \to \frac{2\mu}{\sqrt{1-q}}v^\mu_l = v_{l-1}^{(\mu)}+\eta_lv_{l+1}^{(\mu)} 
\end{equation}
by using a simplified version of the eigen vector
\begin{equation}
    v_l^{(\mu)} = \frac{(1-q)^{l/2}}{(q;q)_l}u_l^{(\mu)}
\end{equation}
where the recursion relation now becomes
\begin{equation}
    2\mu u_l^{(\mu)} = (1-q^l)u_{l-1}^{(\mu)}+u_{l+1}^{(\mu)} \,\, \text{with}\,\, u_{-1}^{(\mu)} =0 \,\, \text{and} \,\, u_{0}^{(\mu)} =1
\end{equation}
this equation defines the recursion relation for Hermite polynomials:
\begin{equation}
    2x H_n(x|q) = (1-q^n)H_{n-1}(x|q)+H_{n+1}(x|q) \,\, \text{with}\,\, H_{-1}(x|q)=0 \,\, \text{and} \,\, H_0(x|q)=1
\end{equation}
Thus the eigen vector is $v_l^{(\mu)} = \frac{(1-q)^{l/2}}{(q;q)_l}H_l(\mu|q)$. 
The matrix T is not hermitian but one can define the symmetric version of it by defining $\hat{T} = P T P^{-1}$ where $P$ is a diagonal matrix with entries $P_l = \frac{\sqrt{(q;q)_l}}{(1-q)^{l/2}}$. This gives the symmetric $\hat{T}_{l_1}^{l_2} = \sqrt{\eta_{l_2}}\delta_{l_1-1}^{l_2}+\sqrt{\eta_{l1}}\delta_{l_1+1}^{l_2} $.
\begin{equation}
\hat{T}=
{\begin{bmatrix}
0 & 1 & 0 & 0 & 0&\cdots\\
1 & 0 & \sqrt{\eta_1} & 0  &0 &\cdots\\
0 & \sqrt{\eta_1} &0& \sqrt{\eta_2}  &0&\cdots\\
\vdots & \ddots &\ddots&&\ddots &\cdots \\
\end{bmatrix}}_{(L+1)\times (L+1)}
\label{symmetrictransfermatrix}
\end{equation}
Now the eigen vectors of the symmetric form of the transfer matrix will take the form
\begin{equation}
    \hat{\psi}_l(\theta) = N(\mu,q)P_l v_l^{(\mu)}= N(\mu,q)\frac{H_l(\mu|q)}{\sqrt{(q;q)_l}} \text{where} \,\, N(\mu,q) = \frac{\sqrt{(q;q)_\infty} |(e^{2 i \theta};q)_\infty|}{\sqrt{2\pi}} 
\end{equation}
now the matrix elements is given by:
\begin{equation}
    \braket{l|\hat{T}^L|m} = \int_0^\pi d\theta  \hat{\psi}_l(\theta) \hat{\psi}_m(\theta) E(\theta)^L
\end{equation}

    General matrix elements of the original Transfer matrix are then
    \begin{equation} \label{eq:lTLm}
        \braket{l|T^L|m} = \int_0^\pi \frac{d\theta}{2\pi}\left( q,e^{\pm2i\theta};q\right)_{\infty }  E(\theta)^L (1-q)^{(l-m)/2} \frac{H_l(x|q) H_m(x|q)}{(q;q)_l}.
    \end{equation}

    One can also compute correlation function of operators $\Psi_I$, with an index set of length $|I| = \tilde{p}$, using the chord diagram prescription. In particular 2-point correlation functions have the simple representation in the dual Hilbert space
\begin{equation}
    \Tr{\mathbf{1} }^{-1} \Tr{\Psi_I e^{\beta_1 H} \Psi_I e^{-\beta_2 H}} = \big< 0 \big| e^{-\beta_1 T}\left( \sum_{k \in \mathbb{N}}  \tilde{q}^k \big| k\big> \big< k \big|\right) e^{-\beta_2 T} \big| 0\big>,
\end{equation}
where $\tilde{q} = e^{-\frac{2 \tilde{p} p}{N}}$ in the double scaled limit, that is when $\tilde{p} \propto \sqrt{N}$.

\section{Double trace correlators in the SYK model} \label{sec:DTinSYK}

We are interested in computing double trace correlators, like the spectral form factor
\begin{equation}
    g(\beta,t) = \Tr{\mathbf{1}}^{-2}\expt{\Tr{e^{-(\beta+i t) H}} \Tr{e^{-(\beta-i t) H}}}. 
\end{equation}
More generally we can consider
\begin{equation}
    G(\beta_1,\beta_2) = \Tr{\mathbf{1}}^{-2}\expt{\Tr{e^{-\beta_1 H}} \Tr{e^{-\beta_2 H}}},
\end{equation}
for arbitrary $\beta_{1,2}$.

To evaluate this double trace quantity, we can use the trick of writing a double trace quantity as a sum over 2-point functions of basis operators, a single trace quantity. The idea is that any double trace quantity looks like:
\begin{equation}
    \Tr{A} \Tr{B} = \sum_{m,n}\bra{m}A\ket{m} \bra{n} B\ket{n} = \sum_{m,n} \Tr{A O_{m,n} B O_{m,n}^\dagger},
    \label{doubletrtosingletr}
\end{equation}
where $O_{m,n} = \ket{m} \bra{n}$. Of course there is nothing special about the basis $O_{m,n}$, and one can work with any orthonormal operator basis $\{O_I\}$ which satisfy $\Tr{O_I^\dagger O_J} = \delta_{I,J}$. In our problem it is natural to work with the basis $\{\Tr{\mathbf{1}}^{-1/2} \Psi_I\}_{I \subset \mathbb{Z}_N}$, which form an orthonormal basis for operators in the Hilbert space.

Thus 
\begin{equation}
    G(\beta_1,\beta_2) = 2^{-N}  \sum_{I\subset \mathbb{Z_N}} \frac{\expt{ \Tr{\Psi_I e^{-\beta_1 H} \Psi_I e^{-\beta_2 H}}}}{\Tr{\mathbf{1}}} .
\label{onetraceonhphysical}
\end{equation}

Assuming there are no triple intersections, we can evaluate this in the double scaled Hilbert space as
\begin{equation} \label{eq:2-pt-fun}
    G(\beta_1,\beta_2) = 2^{-N} \sum_{m = 0}^N \binom{N}{m} \big< 0 \big| e^{-\beta_1 T}\left( \sum_{k \in \mathbb{N}}  \tilde{q_m}^k \big| k\big> \big< k \big|\right) e^{-\beta_2 T} \big| 0\big> .
\end{equation}

While $\tilde{q_m} = e^{-2mp/N}$ is only valid when $p\sim \sqrt{N}$, a more general definition of $\tilde{q}$ was given in appendix F of \cite{Cotler:2016fpe}, where they also derived the crossing factor from first principals. In general for two operators of length $p,m$ the expected value given by their crossing is 
\begin{equation}
    \tilde{q}(p,m;N) = \binom{N}{p}^{-1} \sum_{l=0}^p \binom{m}{l}\binom{N-m}{p-l} (-1)^{l+pm}.
    \label{tildeq}
\end{equation}
Notice that $\tilde{q}(p,m;N) = \tilde{q}(m,p;N)$, and that $q$ is the double scaled limit of $\tilde{q}(p,p;N)$. In fact, using a Q-Gaussian distribution with $q = \tilde{q}(p,p;N)$ is a much better approximation of the spectrum at finite $N$ and $p$ than using the double scaled value $e^{-2p^2/N}$ \cite{Garc_a_Garc_a_2017}.

Furthermore, $\tilde{q}(p,m;N)$ satisfies the identities
\begin{equation} \label{eq:tqstuff}
    \sum_{m=0}^N \binom{N}{m} \tilde{q}(p,m;N) = 0, \qquad \qquad 2^{-N} \sum_{m=0}^N \binom{N}{m} \tilde{q}(p,m;N)^2 = \binom{N}{p}^{-1}.
\end{equation}

As the connected double trace moments contain all contractions of Hamiltonians between the two traces, we can compute the subset of all diagrams where the connected contractions are in parallel, so there are no triple intersections, in a similar fashion to \eqref{eq:2-pt-fun}:
\begin{equation} \label{eq:G_ladder}
    G_{\text{across}}(\beta_1,\beta_2) = 2^{-N} \sum_{m = 0}^N \binom{N}{m} \sum_{k \in \mathbb{N}}  \tilde{q_m}^k \big< k \big| e^{-\beta_1 T} \big| 0\big>  \big< k \big| e^{-\beta_2 T} \big| 0\big> .
\end{equation}
Note that the $T$ is not Hermitian so \eqref{eq:G_ladder} and \eqref{eq:2-pt-fun} are not the same, and that \eqref{eq:G_ladder} is a generalization of the formula in Appendix F of \cite{Cotler:2016fpe}, allowing for additional Hamiltonian insertions.

We can rewrite \eqref{eq:G_ladder} using \eqref{eq:lTLm} and \eqref{eq:intHtoI}, in a similar fashion to equation (C.5) of \cite{Berkooz_2019}, resulting in:
\begin{equation} \label{eq:G_ladder_2}
\begin{aligned}
G_{\text{across}}(\beta+it,\beta-it) =& 2^{-N} \sum_{m = 0}^N \binom{N}{m} \sum_{k,j_+,j_- \in \mathbb{N}} \frac{\tilde{q}^k_m (1-q)^{k+1}}{((q;q)_k)^2(\beta^2 + t^2)} \times \\
& \times \prod_{a=\pm 1} (-1)^{j_a} \frac{q^{j_a(j_a+1)/2}(q;q)_{k+j_a}}{(q;q)_{j_a}} (k+2j_a +1) I_{k+2j_a +1}\left(\frac{2\beta + a 2 i t}{\sqrt{1-q}} \right)     
\end{aligned}
\end{equation}

At late times $t \gg \beta, t \gg \sqrt{1-q} (k+2j + 1)^2  $ each element in the sum decays like $t^{-3}$.  $\tilde{q}_m$ puts an effective cutoff on $k$: $k \lesssim O(N/m)$, so if $t \gg (N/m)^2$ then the whole 2-point function decays as $t^{-3}$. (We assume here that $q$ is of order one and so $j_a$ is also of order 1.) 

While \eqref{eq:G_ladder_2} only captures a subset of diagrams, and ignores triple intersections, it still gives us the necessary intuition for understanding the time-scales and some insight to the emergence of the late time ramp and plateau. Overall \eqref{eq:G_ladder_2} explains the 3 time-frames relevant for the spectral form factor of the SYK model:
\begin{enumerate}
    \item At early times $t\lesssim N^{p/2}$ the sum in \eqref{eq:G_ladder_2} is dominated by small $k$. In this case we can use \eqref{eq:tqstuff} to see that the sum over $m$ is zero for $k=1$ and is dominated by $k=2$. Furthermore, we can evaluate this sum to be
    \begin{equation}
        G_{c,\text{early}}(\beta ,t) = \binom{N}{p}^{-1} \left|\int_0^\pi \frac{d\theta}{\pi} \left(q,e^{\pm 2i\theta} ;q\right)_{\infty}  \frac{e^{(-\beta + it) E(\theta) }}{1+q} \left(E(\theta)^2 - 1 \right) \right|^2.
    \end{equation}
    While it is not quite the correct answer at early times computed by \cite{Berkooz2021}, it still has the correct scaling of $\binom{N}{p}^{-1}$. Furthermore, at times $t \gg \beta$ the connected SFF decays as $t^{-1}$, while the disconnected piece dominates the full spectral form factor.

    \item At intermediate times $N \lesssim t \lesssim 2^{ N/4}$ the sum in \eqref{eq:G_ladder_2} is no longer dominated by small $k$, but rather by $\tilde{q} \approx 1$. Nevertheless, at these time each term in \eqref{eq:G_ladder_2} decays as $t^{-3}$, and we may expect the connected part of the SFF to undergo a transition. This is in fact what was seen numerically in \cite{C_ceres_2022} for the sparse SYK model.

    \item At late times  $ t \gtrsim 2^{N/4}$ the SFF is dominated by the RMT ramp, which is exponentially suppressed in $N$ but grows like $t^1$. \cite{Cotler:2016fpe} argued this corresponds to the element $m=0$, or $\tilde{q} = 1$ in the sum \eqref{eq:G_ladder_2}, as it is the only element that survives at exponentially late times. At even later times  $ t \gtrsim 2^{N/2}$ the SFF transitions to the RMT plateau, which is a constant in time.
    
\end{enumerate}

\begin{figure}
    \centering
    \includegraphics[width=0.6\linewidth]{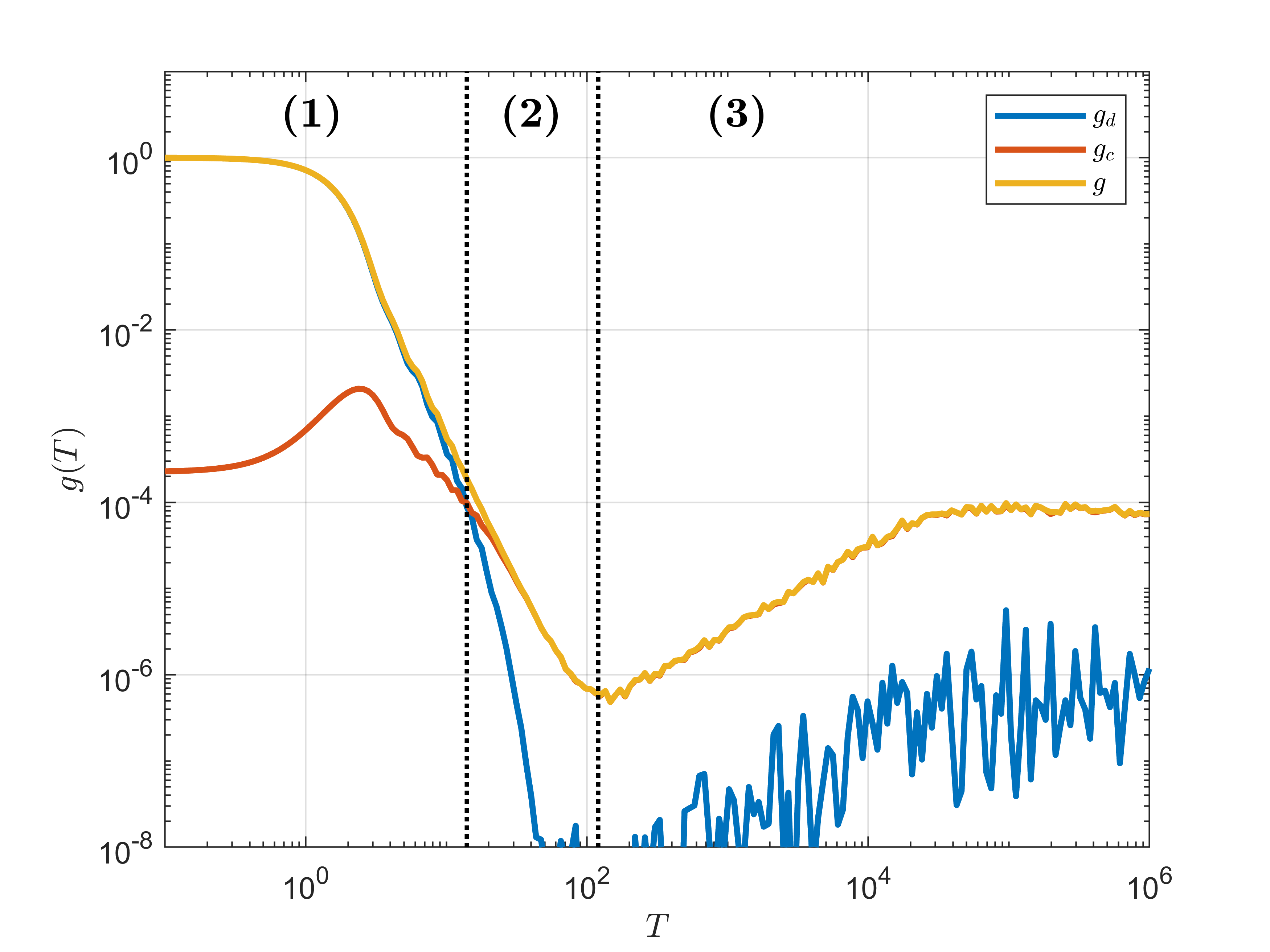}
    \caption{The SFF of the sparse SYK model averaged over 100 realizations, with $N=30$, $p=4$, $\beta = 1$, and sparseness parameter $k=8$, based on data from \cite{C_ceres_2022}. The three distinct timescales are separated by dotted vertical lines, and labeled.}
    \label{fig:timescales}
\end{figure}

To illustrate these three distinct time-frames and behaviors, we include a plot of the numerical SFF in the sparse SYK model in figure \ref{fig:timescales}, where all three regions are clearly seen. We also summarize these results in table \ref{tab:timescales}.

\begin{table}[]
    \centering
    \begin{tabular}{|c||c|c|c|c|} 
        \hline
        label & times & time-frame & dominant piece & time scaling  \\
         \hline \hline
        (1)a & Very Early & $t \lesssim \beta $ & Disconnected &  $g \approx $ constant\\ \hline
        (1)b & Early & $\beta \lesssim t\lesssim N^{p/2}$ & Disconnected &  $g = g_d \propto t^{-3}, ~~g_c \propto t^{-1} $\\ \hline
        (2) & Intermediate & $N^{p/2} \lesssim t \lesssim 2^{ N/4}$& Connected & $g \propto t^{-3}$\\ \hline
        (3)a & Late &  $2^{ N/4} \lesssim t \lesssim 2^{ N/2}$ & Connected & $g \propto t$ \\ \hline
        (3)b & Very late & $t \gtrsim 2^{N/2}$ & Connected & $g \approx $ constant\\ \hline
    \end{tabular}
    \caption{The various time-frames of the SFF. The labels correspond to the time frames seen in figure \ref{fig:timescales}.}
    \label{tab:timescales}
\end{table}

The main interest of this paper will be to access the universal ramp at late times from a chord diagram perspective. \cite{Cotler:2016fpe} argued that, this late time ramp corresponds to  $m=0$ in the sum \eqref{eq:G_ladder_2}, or equivalently to the identity operator in \eqref{onetraceonhphysical}. On it's face this assertion seems odd, as the identity operator contribution in \eqref{onetraceonhphysical} gives something proportional to $Z(\beta_1+\beta_2)$, which is not a ramp. Nevertheless, they argued that these diagrams are the ones that contribute in random matrix theory, based on the ladder computation of \cite{brezin1994correlation}. We will see that this conclusion is a little oversimplified, and that we are in fact missing additional operators that are similar to the identity and also contribute at late times. To show this, we start from analyzing chord diagrams in pure random matrix theory, which is the $q \ra 0$ limit of DSSYK. Then we extend these results to finite $q$ in the following section.

\section{Chords in Random Matrix Theory}
\label{chords in rmt}

In order to compute the late time ramp of the SFF in the DSSYK model, we will start from the random matrix theory limit $q = 0$, and then generalize the results to finite $q$. At $q=0$ DSSYK becomes a Gaussian random matrix theory \cite{Erdos14}. This random matrix theory at large Hilbert space dimension $L = 2^{N/2}$ can be solved in a topological expansion, where the topological genus is $g = L^{-1}$. This topological expansion follows from the t' Hooft double-line notation, where only planar diagrams contribute at leading order, and non-planar diagram are suppressed in powers of the genus \cite{t1993planar,brezin1978planar,brezinzee,brezin1994correlation,brezin1995universal,Saad:2019lba}. In this work we will compute the same quantities using chord diagrams and building an auxiliary Hilbert space, which is somewhat similar to the t' Hooft expansion. 

To illustrate the similarities and differences between these approaches, consider the disk partition function. At leading order only planar diagrams contribute in the genus expansion, while for chord diagrams only the non-crossing ones contribute. In this case the two sets of diagrams are one and the same, and so the results exactly match. At next order in the topological expansion we have diagrams that can be placed on a disk with one handle-body. In terms of chord diagrams, these can have many crossings (see for example figure \ref{connected2},) so the topological expansion is not strictly an expansion in the number of crossings.\footnote{In particular all diagrams with one crossing can be placed on a disk with one handle, and so are suppressed by $g^{-2}$. However, some diagrams with two crossing can be put on a disk with one handle, while other diagrams require two handles.} Nevertheless, we can exactly map the t' Hooft topological expansion to non-crossing chord diagrams on various topologies weighted by the genus.

 To be concrete, we consider a $L \times L$ random Hermitian matrix $H$ drawn from a Gaussian ensemble. The moments $m_k = \expt{\Tr{H^k}}$ can be computed using chord diagrams similar to the DSSYK model in the $q \ra 0$ limit. The only diagrams that contribute in the $L \ra \infty$ limit are planar diagrams, which necessarily have no crossings. Thus we can compute the moments in the auxiliary Hilbert space as 
\begin{equation}
    m_k = \expt{\Tr{H^k}} = \left< 0 | T^k |0 \right>,
\end{equation}
with the transfer matrix $T = a + a^\dagger$, and the raising and lowering operators $a \ket{n} = \ket{n-1}$, $a^\dagger \ket{n} = \ket{n+1}$. As $T$ is a tri-diagonal matrix with ones above and bellow the diagonal. The eigenvalues of this transfer matrix are $E(\theta) = 2 \cos(\theta)$, $\theta \in [0,\pi]$, with corresponding eigenvectors $\ket{\theta} = \sum_n v_n \ket{n}$  whose coefficients satisfy the recurrence relation
\begin{equation}
    2 \cos(\theta) v_n = v_{n-1} + v_{n+1}, \qquad v_{-1} = 0, ~~v_0 = A. 
\end{equation}
The solution to this recurance relation tells us that 
\begin{equation}
    v_n = A(\theta) U_n(\cos \theta),
\end{equation} 
where $ U_n (x)$ is the $n$'th Chebyshev polynomials of the second kind, and $A$ is a normalization constant. We can compute $A$ using the completeness relation of Chebyshev polynomials:
\begin{equation}
    \left< \theta | \phi \right> = A(\theta) A(\phi) \sum_n U_n(\cos \theta) U_n(\cos \phi) = A(\theta) A(\phi) \frac{\pi}{2\sqrt{|\sin\theta \sin \phi|}} \delta(\cos \theta - \cos \phi) .
\end{equation}
Thus $A(\theta) = \sqrt{2/\pi} |\sin \theta|$ so that the eigenvectors form a complete basis with $\delta$-function normalization $\left< \theta | \phi \right> = \delta(\theta - \phi)$.

The moments of the Hamiltonian are then 
\begin{equation}
    m_k = \left< 0 | T^k |0 \right> = \int_0^\pi d\theta \frac{2 }{\pi} \sin^2(\theta) [2\cos(\theta)]^k = \left\{ \begin{array}{cc}
        C_{k/2}, &  k \in 2\mathbb{N}, \\
        0, & k \in 2\mathbb{N} +1.
    \end{array} \right.
    \label{eq:singletrRMT}
\end{equation}
where $C_{k/2}$ are the Catalan numbers. These are the expected moments for a large random (properly normalized) Hermitian matrix \cite{anderson2010introduction}. The density of eigenvalues can also be read off as $\rho(\theta) = \frac{2}{\pi}\sin^2(\theta)$, which is just the semicircle law for the energies $E = 2\cos(\theta)$.

\begin{figure}
    \centering  \includegraphics[width=0.4\linewidth]{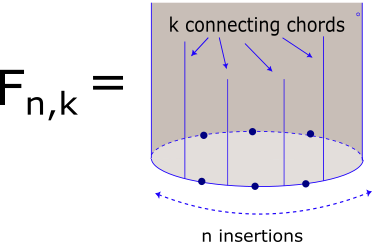}
    \caption{$F_{n,k}$ counts every possible configuration of planar diagrams of a single boundary with $n$ Hamiltonian insertions, $k$ of which connect to the other trace.}
    \label{F_nk}
\end{figure}

For double trace moments the computation is slightly more involved. We want to compute the connected moments
\begin{equation}
    m^c_{n_1,n_2} = \expt{\Tr{H^{n_1}} \Tr{H^{n_2}}}_c .
\end{equation}
We know that these will be given by summing over planar diagrams with two boundaries. To get the connected piece we can sum over the number of Hamiltonians connecting the two traces, which we label by the integer $k$. Then we can write the moments as
\begin{equation}
        m^c_{n_1,n_2} = \frac{1}{L^2} \sum_{k\geq 1} k F_{n_1,k} F_{n_2,k},
\end{equation}
where $F_{n,k}$ is the number of planar diagrams of a single boundary with $n$ Hamiltonian insertions, $k$ of which connect to the other trace. See Figure \ref{F_nk}. The factor $k$ arises from the $k$ cyclic ways to connect the two traces given that the connecting chords are chosen.

From a topological viewpoint, $F_{n,k}$ is a trumpet partition function where $n$ corresponds to the boundary state and $k$ is discrete length of the interior geodesic, similar to \cite{Saad:2019lba,okuyama2023endworldbranedouble}. In this context it is natural to define the trumpet partition function as $F_{\beta,k} = \sum_{n \geq 0} \frac{(-\beta)^n}{n!} F_{n,k}$, where $\beta$ is the length of the boundary. Then the cylinder partition function $G(\beta_1,\beta_2)= \sum_{k\geq 1} k F_{\beta_1,k} F_{\beta_2,k}$ can be thought of as a sum of the discrete ways to glue the two trumpets to make a cylinder.

While we might think we can simply compute $F_{n,k}$ in the chord space as $F_{n,k} = \left< 0 \mid T^n\mid k \right>$, this unfortunately doesn't count all the planar diagrams as some diagrams can have chords wrapping around the trace, resulting in apparent crossings (see Appendix A-E). We can account for this by adding a "sea" of say $l > n-k$ open chords, which may or may not close during the Hamiltonian evolution, and compute $F_{n,k} =\left< l \mid T^n\mid k+l \right>$. This "sea" of chords allows for chords to be closed before they are opened, and counts all the planar chord diagrams for large enough $l$. As we would like the auxiliary Hilbert space to be the same for all computation of the moments, we need to take $l \ra \infty $, or equivalently work in the Hilbert space which allows "negative" chord number states $\mathcal{H} = \{\ket{n} \}_{n\in \mathbb{Z}}$.

We need to now diagonalize $T$ on this new space. Fortunately, $T$ will have the same spectrum of Eigenvalues and the eigenvectors will have the same recurrence relation, only now the initial condition $v_{-1} = 0$ is no longer valid. This will result in two independent eigenvector for each eigenvalue $E(\theta) = 2\cos(\theta)$, which we can take to be 
\begin{equation}\begin{aligned}
    \ket{\theta,1} &= \frac{1}{\sqrt{\pi}} \sin(\theta) \left( \sum_{n\geq 1 } U_{n-1}(\cos \theta) \ket{n} - \sum_{n\leq -1 } U_{|n|-1}(\cos \theta) \ket{n}\right) =  \frac{1}{\sqrt{\pi}} \sum_{n\in \mathbb{Z}} \sin(n\theta) \ket{n},\\
    \ket{\theta,2} &= \frac{1}{\sqrt{\pi}} \left( \sum_{n\geq 0 } T_{n}(\cos \theta) \ket{n} + \sum_{n<0 } T_{|n|}(\cos \theta) \ket{n}\right) =  \frac{1}{\sqrt{\pi}} \sum_{n\in \mathbb{Z}} \cos(n\theta) \ket{n}.
    \end{aligned}
\end{equation}
It is easy to see that $\left< \theta,1|\theta' ,2 \right> = 0$, as $\ket{\theta,1}$ is anti-symmetric while $\ket{\theta,2}$ is symmetric in the $\ket{n}$ basis. One can also check that they are properly normalized as $\left< \theta,i|\theta' j \right> = \delta_{i,j} \delta(\theta - \theta')$ using the sum representation $2\pi \delta(\theta) = \sum_{n \in \mathbb{Z}} e^{i n \theta} $.

Now we can compute $F_{n,k}$ as
\begin{equation} \label{eq:F_qzero}
    F_{n,k} = \left<k \mid T^n \mid 0\right>
    = \frac{1}{\pi} \int_0^\pi d\theta (2\cos \theta)^n \cos(k\theta),
\end{equation}
where the inner product here is in the modified Hilbert space. Then we can sum these to get the connected moments
\begin{equation}
\begin{aligned}
    m^c_{n_1,n_2} &= \frac{1}{L^2} \frac{1}{\pi^2} \int_0^\pi d\theta_1 d\theta_2  (2\cos \theta_1)^{n_1}  (2\cos \theta_2)^{n_2}\sum_{k\geq 1} k \cos(k \theta_1) \cos(k \theta_2)\\
    &= \frac{1}{L^2} \frac{1}{\pi^2} \int_0^\pi d\theta_1 d\theta_2  (2\cos \theta_1)^{n_1}  (2\cos \theta_2)^{n_2}\frac{\cos(\theta_1) \cos(\theta_2)-1}{2\left(\cos(\theta_1) - \cos(\theta_2) \right)^2}.
\end{aligned}
\end{equation}
We can rewrite this in terms of the energies $E_i = 2\cos \theta_i$ as
\begin{equation}
\begin{aligned}
    m^c_{n_1,n_2} &= \frac{1}{L^2} \frac{1}{\pi^2} \int_{-1}^1 \frac{dE_1 dE_2}{\sqrt{4-E_1^2} \sqrt{4-E_2^2}}  (E_1)^{n_1}  (E_2)^{n_2} \frac{E_1 E_2-4}{2\left(E_1 - E_2 \right)^2},
\end{aligned}
\end{equation}
leading to the eigenvalue correlation
\begin{equation}
    \rho_c(E_1,E_2) = \frac{1}{L^2 \pi^2} \frac{1}{\sqrt{(4-E_1^2) (4-E_2^2)}}\frac{E_1 E_2-4}{2\left(E_1 - E_2 \right)^2}.
    \label{eq:doubletrRMT}
\end{equation}
This is precisely the universal eigenvalue correlation function of Brezin and Zee \cite{brezinzee}, and matches the results from topological recursion \cite{Saad:2019lba}.\footnote{More precisely, \cite{Saad:2019lba} compute the universal resolvent $R_{0,2}(E_1,E_2)$. The spectral correlation function is related to this resolvent by \cite{brezinzee} \[
L^2 \rho_c(E_1,E_2) = -\frac{1}{4\pi^2}\left( R_{0,2}(E_1+i\epsilon,E_2+i\epsilon) + R_{0,2}(E_1-i\epsilon,E_2-i\epsilon) - R_{0,2}(E_1-i\epsilon,E_2+i\epsilon) - R_{0,2}(E_1+i\epsilon,E_2-i\epsilon)\right).  \]}

We can also compute the double-trace partition function. To do this, it is simplest to start with the trumpet partition function
\begin{equation} \label{eq:Fbkq0}
    F_{\beta,k} = \frac{1}{\pi} \int_0^\pi d\theta e^{- 2\beta \cos \theta } \cos(k\theta) = (-1)^k I_{k}(2\beta),
\end{equation}
where $I_k(x)$ is the modified Bessel function of the first kind. Then the double-trace partition function is \cite{okuyama2023endworldbranedouble}
\begin{equation}
\begin{aligned}
    G(\beta_1,\beta_2) &=\frac{1}{L^{2}} \sum_{k \geq 1} k  F_{\beta_1,k}  F_{\beta_2,k} = \frac{1}{L^{2}}\sum_{k \geq 1} k I_{k}(2\beta_1) I_{k}(2\beta_2)\\
    &= \frac{1}{L^{2}} \frac{\beta_1 \beta_2}{\beta_1 + \beta_2} \left( I_0(2\beta_1) I_1(2\beta_2) + I_1(2\beta_1) I_0(2\beta_2) \right).
\end{aligned}
\end{equation}

\section{From RMT to finite \texorpdfstring{$q$}{q}}
\label{fromrmttofiniteq}
\begin{figure}
\centering   
\includegraphics[width=0.2\linewidth]{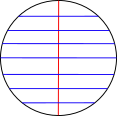}
\qquad
\includegraphics[width=0.4\linewidth]{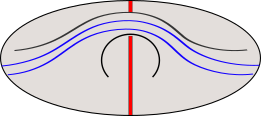}
    \caption{A contribution to the leading order connected double-trace moments in RMT where some of the Hamiltonian chords (blue) cross the marked chord (red), but no crossings exist between Hamiltonian chords. One can also view these diagrams as non-crossing chord diagrams in the topology of a disk with a single handle-body. Due to the crossing, the marked chord must be the identity operator to ensure a non-vanishing contribution.}
    \label{connected1}
\end{figure}

\begin{figure}
\centering
\includegraphics[width=.8\linewidth]{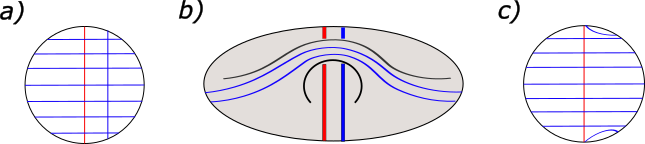}
\caption{a) A contribution to the connected double-trace moments in RMT where the Hamiltonian chords that cross the marked chord also intersect a different Hamiltonian chord. b) One can again view these diagrams as non-crossing chord diagrams in the topology of a disk with a single handle-body, where the intersecting Hamiltonian chord runs parallel to the marked chord through the handle-body. c) Due to the crossings, the marked chord should be identified with the parallel Hamiltonian chord to ensure a non-vanishing contribution.}
    \label{connected2}
\end{figure}

\begin{figure}
\centering
\includegraphics[width=0.8\linewidth]{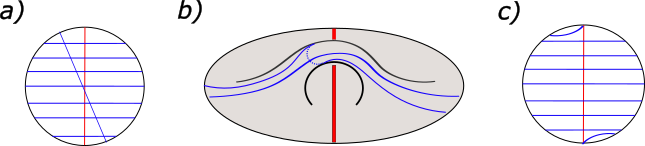}
\caption{a) A contribution to the connected double-trace moments in RMT where the Hamiltonian chords that cross the marked chord also intersect each other. b) One can again view these diagrams as non-crossing chord diagrams in the topology of a disk with a single handle-body, where the intersecting Hamiltonian chord twist around the handle. c) Due to the crossings, the marked chord should be identified with the twisting Hamiltonian chord to ensure a non-vanishing contribution.}
    \label{connected3}
\end{figure}

We would like to map the computation of the spectral form factor in the RMT limit to a chord interpretation, which we can then make precise at finite $q$. The main insight is to take the representation of the SFF as a sum of 2-point functions seriously, and to argue that only one of $2^N$ operators contributes to each of the relevant diagrams. The relevant non-planar diagrams at leading order can have 3 types of chord intersections: 
\begin{enumerate}
    \item In each diagram some Hamiltonian chords will cross the marked chord. A diagram with only such crossing is presented in figure \ref{connected1} .

    \item There may be chords that run parallel to the marked chord and intersect all Hamiltonian chords that cross the marked chord. A diagram depicting this is presented in figure \ref{connected2}.

    \item The Hamiltonian chords that cross the marked chord may intersect among themselves due to a twist. A diagram depicting this is presented in figure \ref{connected3} .
\end{enumerate}

All these diagrams can also be thought of as non-intersecting chord diagrams on a disk with one handle-body, where the marked chord runs through the hole in the handle-body.

To understand which operator contributes for each diagram, we can start with diagrams with only the first type of intersections. For these diagrams the only way to get a non-vanishing contribution is for the marked chord to be the identity operator, as operators of any other length are suppressed by the intersections. All these diagrams are weighted by $2^{-N} = 1/L^2$ as only one of the $2^N$ operators contributes. These are precisely the diagrams considered in \eqref{eq:G_ladder}. 

The second types of diagram, in the topological picture, has chords running through the handle-body parallel to the marked chord. In the operator expansion we can think of them as a contraction between the Hamiltonian and the marked chord, leading to a non-crossing diagram weighted by $2^{-N} = 1/L^2$ as only one of the $2^N$ operators, namely $\mathcal{O} = I H $ contributes.\footnote{Note that when more than two Hamiltonian chords contract with the marked chord, we must remember to transpose one of the operators, or equivalently reverse the order of the contracting index sets, as per \eqref{doubletrtosingletr}. This ensures the resulting diagram has no intersections.}

The third types of diagram, in the topological picture, has chords twisting as they go through the handelebody. In the operator expansion we can also think of them as a contraction between the Hamiltonian and the marked chord, though with the Hamiltonian chord now on either side of the marked chord. This also generates to a non-crossing diagram weighted by $2^{-N} = 1/L^2$ as only one of the $2^N$ operators, namely $\mathcal{O} = I H $ contributes.

\begin{figure}
\centering
\includegraphics[width=1\linewidth]{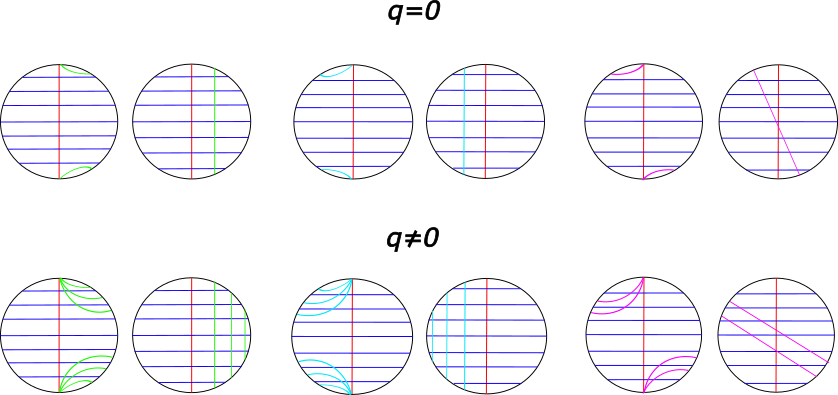}
    \caption{For $q=0$, we have 3 sets of chords. We allow only global intersections when twisting because this will amount to no intersections at all in the marked chord expansion around Hamiltonian insertions as you can see on the left diagram of each set. For the finite q, we again have 3 sets of diagrams. We allow for local intersections like the right side chord diagram where two crossing chords twist and therefore intersect among the other chords crossing between the two traces as one can see on the left diagram of the right set. Notice that global intersections here are subleading as $q^{\# \text{large number}}$ is heavily suppressed so it won't contribute at leading order. However, a small number of intersections (in other words, local intersection) can contribute when $q$ is small at leading order in $2^{-N}$, and should be accounted for.}
    \label{fig:cdforq=0&finiteq}
\end{figure}

Overall, if we wanted to evaluate this at finite $q$ we should expand the marked chord around different insertions of the Hamiltonian, and possibly allow for a small operator as well. While at early times the long operators will contribute, and so this will be an over-counting, at late times only the identity in each channel remains, and so these will be the precise diagrams that contribute, and lead to the universal ramp. To illustrate the difference between the $q\to0$ diagrams and those of finite $q$, we collect the relevant diagrams in figure \ref{fig:cdforq=0&finiteq}.

To compute these contributions, we note that the suppression factor will be the same $2^{-N} = 1/L^2$, as a single operator is contributing. Then we can consider all diagrams with $l_{1,2}$ chords of type 2 on the left/right trace, $r$ chords of type 3 that twist, and $s$ remaining chords that connect the two traces. Taking into account the allowed intersections between the different types of chords we have that the full late time double-trace partition function is 
\begin{equation} \label{eq:Glatedef}
    \begin{aligned}
        G_{\text{late}}(\beta_1,\beta_2) &= 2^{-N} \sum_{l_1,l_2,r \geq 0, ~s>0} \left<l_2 \mid \vdash e^{-\beta_2 T} \dashv \mid r+s+l_2\right> {s+r+l_1 \brack s,r,l_1}_q \left<s+r+l_1 \mid \vdash e^{-\beta_1 T} \dashv \mid l_1\right> ,
    \end{aligned}
\end{equation}
where $\left<n\mid \vdash e^{-\beta T} \dashv \mid m\right>$ is the sum over all chord diagrams with $m$ initial chords and $n$ final chords and evolution $e^{-\beta T}$, such that all initial and final chords close on in the interval. The coefficient ${s+r+l \brack s,r,l_1}_q = \frac{(q;q)_{s+r+l}}{(q;q)_s (q;q)_r (q;q)_l}$ is the $q$-mutinomial coefficient, and accounts for the intersection factor given by the choice of which of the $s+r+l_1$ chords belong to each group.

\begin{figure}
\centering
\includegraphics[width=.7\linewidth]{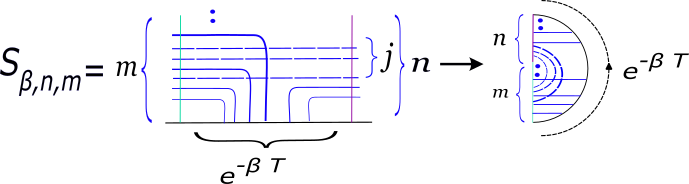}
\\
\includegraphics[width=.7\linewidth]{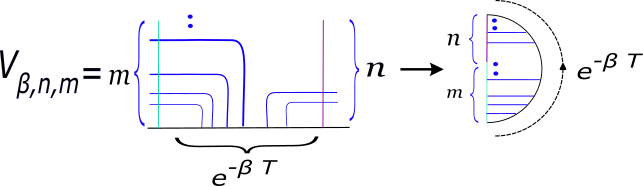}
    \caption{An illustration of $S_{\beta,n,m}$ and $V_{\beta,n,m}$. In $S_{\beta,n,m}$, we have chords that don't close on the boundary. We are interested in the diagrams that don't have open chords therefore we define $V_{\beta,n,m}$ as those diagrams in S that don't have open chords. from the definition of $V_{\beta,n,m}$, one can draw the types of diagrams that will contribute to G late}
    \label{fig:sandv}
\end{figure}

To compute these modified inner products, we need to relate them to the standard inner products in the auxiliary Hilbert space. Using the notation
\begin{equation}
    S_{\beta,n,m} \equiv \left<n\mid  e^{-\beta T}  \mid m\right>, \qquad \qquad V_{\beta,n,m} \equiv \left<n\mid \vdash e^{-\beta_1 T} \dashv \mid m\right>, 
    \label{sandv}
\end{equation}
we would like to relate the two quantities. While $S_{\beta,n,m}$, contains all the diagrams in $V_{\beta,n,m}$ this would be an overcounting as $S_{\beta,n,m}$ also contains diagrams where some of the $m$ initial chords don't close. So, we introduce $V_{\beta,n,m}$ where all chords close after the Hamiltonian evolution, hence the notation $\dashv$. However, we can relate them by
\begin{equation} \label{eq:StoV}
    S_{\beta,n,m} = \sum_{j=0}^{\min\{m,n\}} {m \brack j}_q V_{\beta,n-j,m-j} ,
\end{equation}
where ${n \brack k}_q = \frac{(q;q)_n}{(q,q)_{n-k}(q,q)_{k}} $ is the $q$-binomial coefficient. The $q$-binomial coefficient comes from the possible choice of which of the $m$ initial chords do not close, giving a factor due to possible intersections. For example, if all of the initial $m$ chords chords close during the Hamiltonian evolution then we have simply $V_{\beta,n,m}$. Now if one of the initial chords does not close, then we have $m$ choices of chords. For the $j$'th chord Hamiltonian evolution will give the factor $q^{j-1} V_{\beta,n-1,m-1}$, where the factor of $q^{j-1}$ came from closing all the chords above the $j$'s chord. Then the full prefactor of $V_{\beta,m-1,n-1}$ in $S_{\beta,n,m}$ will be $1 + q + \ldots +q^{m-1} = \frac{1-q^m}{1-q} = {m \brack 1}_q$. We can compute the other prefactors in a similar manner. See Figure \ref{fig:sandv}. 
Using the analytic evaluation of the inner product, \eqref{eq:lTLm}, along with the linearizing relation of the $q$-Hermite polynomials, \eqref{eq:HHtoH}, we find that
\begin{equation}
\begin{aligned}
    S_{\beta,n,m} &=  \int_0^{\pi} \frac{d\theta}{2\pi} \left(q,e^{\pm 2i\theta};q\right)_\infty e^{-\beta E(\theta)} (1-q)^{(n-m)/2} \frac{H_{n}(x|q)H_{m}(x|q)}{(q;q)_{n}}\\
    &= \int_0^{\pi} \frac{d\theta}{2\pi} \left(q,e^{\pm 2i\theta};q\right)_\infty e^{-\beta E(\theta)} (1-q)^{(n-m)/2} \sum_{j=0}^{\min\{n,m\}} {m \brack j}_q \frac{H_{n+m-2j}(x|q)}{(q;q)_{n-j}} .
\end{aligned}
\end{equation}
By direct comparison to \eqref{eq:StoV}, we see that
\begin{equation}
    V_{\beta,n,m} = \int_0^{\pi} \frac{d\theta}{2\pi} \left(q,e^{\pm 2i\theta};q\right)_\infty e^{-\beta E(\theta)} (1-q)^{(n-m)/2} \frac{H_{n+m}(x|q)}{(q;q)_{n}} .
\end{equation}

Then, plugging this into \eqref{eq:Glatedef}, the topological part of the double-trace partition function is
\begin{equation} \label{eq:Glatesum}
    \begin{aligned}
        G_{\text{late}}(\beta_1,\beta_2) &= 2^{-N}\int_0^\pi \frac{d\theta_1 d\theta_2}{(2\pi)^2} \left(q,q,e^{\pm 2i\theta_1},e^{\pm 2i\theta_2};q\right)_\infty e^{-\beta_1 E(\theta_1) -\beta_2 E(\theta_2)} \\
        &\qquad  \qquad \times
        \sum_{l_1,l_2,r \geq 0, ~s>0}  \frac{H_{s+r+2l_1}(x_1|q) H_{s+r+2l_2} (x_2|q)}{(q;q)_s (q;q)_r (q;q)_{l_1} (q;q)_{l_2}} .
    \end{aligned}
\end{equation}
Note that $G_{late}(\beta_1,\beta_2;q) = G_{late}(\beta_2,\beta_1;q)$, as required by symmetry.

\begin{figure}
    \centering
    \includegraphics[width=0.95\linewidth]{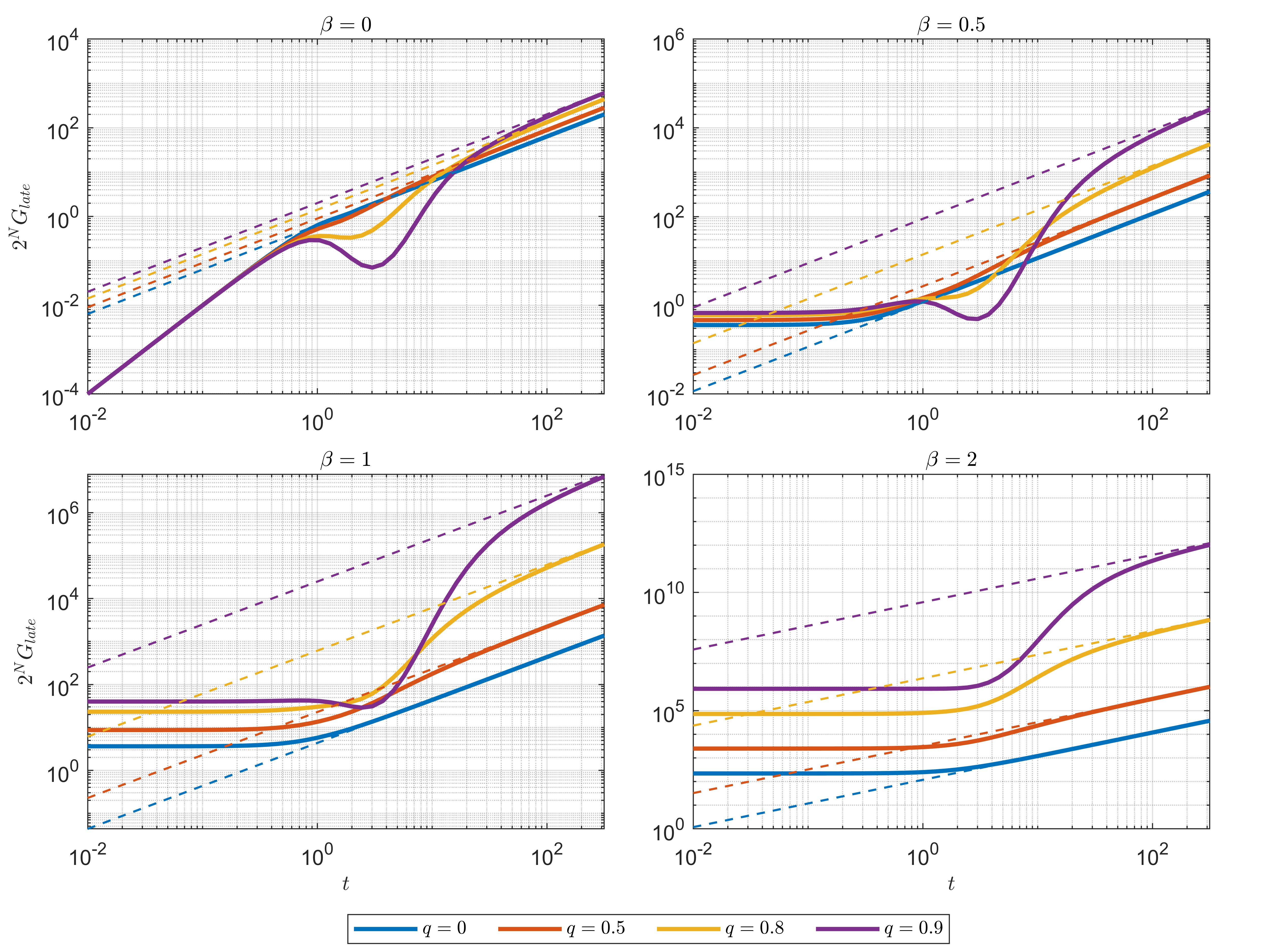}
    \caption{A plot of $G_{\text{late}}(\beta,t;q)$ for various values of $\beta$ and $q$. The solid lines are  $G_{\text{late}}(\beta,t;q)$ given by $\eqref{eq:Glate}$, while the dashed lines are the analytical asymptotes of the ramp given by \eqref{eq:actualramp}.}
    \label{fig:glate}
\end{figure}

We would like to have some topological interpretation of \eqref{eq:Glatesum}, similar to writing the RMT cylinder partition function as a sum over the possible ways of gluing two trumpets. To do that, we can change variables to $k = s+r$ and write
\begin{equation} \label{eq:Glate}
    \begin{aligned}
        G_{\text{late}}(\beta_1,\beta_2) &= 2^{-N} \sum_{k \geq 1} C_k~ F_{\beta_1,k}(q) F_{\beta_2,k}(q),\\
        F_{\beta,k}(q) &= \int_0^\pi \frac{d\theta}{2\pi} \left(q,e^{\pm 2i\theta};q\right)_\infty e^{-\beta E(\theta)} \sum_{l \geq 0} \frac{H_{k+2l}(x|q)}{(q;q)_l},\\
        C_k &= \frac{1}{(q;q)_k}\sum_{r = 0}^{k-1} {k \brack r} = \frac{H_k(1|q) - 1}{(q;q)_k} .
    \end{aligned}
\end{equation}
Here $F_{\beta,k}(q)$ is the trumpet partition function with boundary $\beta$ and inner geodesic $k$ at finite $q$, and the factor $C_k$ is the possible ways of gluing two trumpets of inner geodesics $k$. We can also write the trumpet partition function as a sum of modified Bessel functions (see appendix \ref{app:glate} for the full computation)
\begin{equation} \label{eq:FBes}
    F_{\beta,k}(q) =\sum_{m=0}^\infty  \left(q^{m+1};q\right)_{k-1} (-1)^m q^{\binom{m}{2} + m (k+m)} \left(1-q^{k+2m} \right) I_{k+2m}\left(\frac{2\beta}{\sqrt{1-q}} \right) .
\end{equation}
It is easy to verify that $C_k \ra k$ as $q \ra 0$, while $F_{\beta,k}(q)$ converges to \eqref{eq:Fbkq0} in this limit. 
This sum is easy to plot using standard software, and is presented in figure \ref{fig:glate}.

The joint eigenvalue distribution is
\begin{equation} \label{eq:rholate}
\begin{aligned}
    \rho_{late}(\theta_1,\theta_2;q) &= 2^{-N} \frac{\left(q,q,e^{\pm 2i\theta_1},e^{\pm 2i\theta_2};q\right)_\infty}{(2\pi)^2} \sum_{r,l,j=0, s>0}^\infty  \frac{H_{r+s+2l_1}(x_1|q) H_{r+s+2l_2}(x_2|q)}{(q;q)_{r}(q;q)_{s} (q;q)_{l_1} (q;q)_{l_2}}.
\end{aligned}
\end{equation}

Equations \eqref{eq:Glate}, \eqref{eq:FBes} and \eqref{eq:rholate} are the main results of our computation.

While simplifying \eqref{eq:rholate} is difficult, we can take the limit $\theta_1 \ra \theta_2$, which should contain the universal behavior encoding the RMT ramp \cite{Cotler:2016fpe}. We preform the full computation in appendix \ref{sec:Glate}, and will simply state the result here. Calling $\Delta \theta \equiv \theta_1-\theta_2 \ll 1$, and $\bar{\theta} \equiv (\theta_1 + \theta_2)/2 \approx \theta_1 \approx \theta_2$, we have using the above approximations
\begin{equation}
    \begin{aligned}
        \rho_{late}(\bar{\theta},\Delta\theta;q) & \approx -  \frac{2^{-N}}{8\pi^2 \sin^2(\Delta \theta /2)} .
        \label{eq:asymrholate}
    \end{aligned}
\end{equation}

This limit of the joint eigenvalue distribution is  responsible for the late time ramp in the SFF \cite{Cotler:2016fpe}, and results in the late-time ramp\footnote{See appendix \ref{sec:Glate} and in particular \eqref{eq:glateramp} for the full computation.}
\begin{equation} \label{eq:actualramp}
    G_{\text{late}}(\beta,t \gg 1, q) = 2^{-N} \frac{t}{2\pi} \int_{-E_0}^{E_0} dE ~e^{-2 \beta E} ,
\end{equation}
where $E_0 = \frac{2}{\sqrt{1-q}} = E(\theta=0)$ is the edge of the eigenvalue distribution.

\section{Summary and outlook}
The motivation behind this article is to study the universal ramp in the spectral form factor from the point of view of chord diagrams. Our starting point was writing the spectral form factor, a double trace correlator, as a sum of single trave 2-point functions over a basis of operators, as in \eqref{doubletrtosingletr}. Then, taking the RMT or $q \ra 0$ limit, we found that the non-crossing chord diagrams don't account for all possible leading order diagrams captured by the double line notation in random matrix theory. This is because naively counting non-intersecting diagrams in the $q\to 0$ limit only accounts for the identity operator, and so won’t capture the full topology of the cylinder. In this article, we provided a bridge between RMT and DSSYK by accounting for the missing operators that contribute to the late time ramp,

To find the missing operators, in section \ref{chords in rmt} we computed the well known spectral correlations of RMT using an auxiliary chord space. While the standard chord Hilbert space where chords are not allowed to intersect is sufficient to reproduce single-trace quantities like the semicircle spectrum, as has been well known \cite{Erdos14,Berkooz_2019,feng2018spectrum}; to reproduce the known double-trace correlations one need to use a bigger Hilbert space with negative chord numbers, or a sea of chords. This sea of chords accounts for chords that wrap around the cylinder, and along with a gluing factor, allows us to match the known RMT for spectral correlations, \eqref{eq:doubletrRMT}, \cite{brezinzee,RMTconcepts}.

Starting from the RMT result in terms of a chord Hilbert space, it is simple to map it back to the operator picture: the sea of chords from the modified Hilbert space, along with the gluing factor from twisting chords, are really just operator insertions that are proportional to, and contract with the Hamiltonian. We show this exact correspondence at the beginning of section \ref{fromrmttofiniteq}, where the leading non-planar diagrams are mapped to non-intersecting diagrams with chords that can contract with he marked operator, as shown in figures \ref{connected1}, \ref{connected2}, and \ref{connected3}.

Once we have identified the relevant operators that lead to the late time ramp, it is simple to extend the computation to finite $q$ by allowing chords to intersect, as in figure \ref{fig:cdforq=0&finiteq}. This results in the late-time spectral form factor \eqref{eq:Glatesum}. This sum has a topological interpretation of summing over the possible discrete ways of gluing two trumpets, as shown in \eqref{eq:Glate}. Crucially, both the trumpet partition function and the gluing factor are modified from the RMT limit, though the late time limit of the SFF still produces the universal RMT ramp. Figure \ref{fig:glate} depicts this exact result for $G_{late}$. We also analytically compute the joint eigenvalue distribution in the coincident limit, \eqref{eq:asymrholate}, and verify it reproduces the known RMT ramp at late times, \eqref{eq:actualramp}. This provides an analytical reproduction of the universal ramp in the SFF directly from chord diagrams in the DSSYK.

\begin{figure}
    \centering \includegraphics[width=0.4\linewidth]{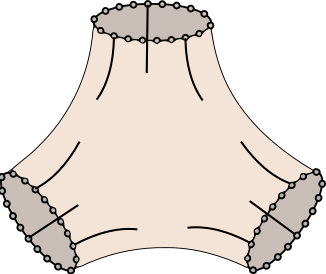}
    \caption{The chord analog to the topological pair of pants.}
    \label{fig:pariofpants}
\end{figure}

\emph{\textbf{Outlook:}} There are several interesting points that arise from our computation, including:
\begin{enumerate}
    \item While the Schwarzian, or $q \ra 1$, limit of DSSYK matches various aspects of JT gravity, and so should be dual to a gravitational theory in AdS$_2$ with matter, the RMT, or $q \ra 0$, limit of DSSYK is a pure gaussian random matrix theory, and so should be dual to a purely gravitational theory with no matter. This is also seen in the double-trace correlators. The finite $q$ DSSYK model contains large global fluctuations to the spectrum which dominate the SFF at early times, which can be thought of as arising from boundary correlations of the matter fields \cite{Berkooz2021}. On the other hand, pure RMT only contains the topological pieces of the SFF, as expected for a purely gravitational theory. Thus, to understand the gravitational dual of DSSYK, and the SYK model in general, it may be simpler to start from the gravitational dual of a Gaussian RMT, and deforming it in a way that reproduces the DSSYK. 

    \item It would be interesting to understand the  Schwarzian limit of $G_\text{late}$, \eqref{eq:Glate}. The Schwarzian limit involves taking $q\ra 1$, while also focusing on the edge of the spectrum. This can be done by scaling $\beta$ as a power of $\lambda^{-1}$, where $q^{-\lambda}$ and we are taking $\lambda \ra 0$ \cite{Micha2018}. Based on figure \ref{fig:glate}, it seems like the SFF behaves similar to the universal RMT SFF in this limit, though making this observation more precise would be interesting.

    \item The combinatorics of chord diagrams developed here can be extended to a full picture of topological recursion for chord diagrams. While we computed the analogs of the trumpet partition function and the gluing factor, we can also consider a chord version of the pair of pants as in figure \ref{fig:pariofpants}. Together these can be used to construct a chord versions of topological recursion, in analogy to \cite{Saad:2019lba}. A similar problem also arises in quiver diagrams \cite{michatalk}.

\end{enumerate}

\section*{Acknowledgment}
We would like to thank Micha Berkooz, Vladimir Narovlansky,  and Joaquin Turiaci for fruitful discussions. This work was supported, in part, by the U.S. Department of Energy under
Grant DE-SC0022021 and by a grant from the Simons Foundation (Grant
651678, Andreas Karch).
\appendix

\section{Special Functions}

There are many $q$-analogs of various functions used throughout our computations. We will provide the general definitions and properties of interest to us, and refer the readers to \cite{HypergeometricSeries,Hypergeometric_Book} (or other standard textbooks) for further details. Throughout this appendix we will assume $|q| \leq 1$. 

The $q$-Pochhammer symbol is defined as 
\begin{equation} \label{eq:qpochammer}
    \left(a;q\right)_n = \prod_{k=0}^{n-1} \left(1 - a q^k \right).
\end{equation}
This definition can be extended to non-integer $n$ by
\begin{equation}\label{eq:qpochgen}
    \left(a;q\right)_\lambda = \frac{(a;q)_\infty}{(a q^\lambda ;q)_{\infty}}.
\end{equation}
The $q$-Pochhammer symbol satisfies many properties, including \cite{Hypergeometric_Book}
\begin{equation} \label{eq:qpochprops}
    \begin{aligned}
        (a;q)_n &= \left(1-aq^{n-1}\right)(a;q)_{n-1},\\
        (a;q)_{n+k} &= \left(a;q\right)_n \left(a q^n;q\right)_k ,\\
        \left(q^{-n};q\right)_k &= \frac{(q;q)_n}{(q;q)_{n-k}}(-1)^k q^{\binom{k}{2} - kn}.
    \end{aligned}
\end{equation}

We will also use the standard notation \cite{Micha2018}
\begin{equation}
    \left(a_1,a_2,\ldots a_m;q\right)_n = \prod_{k=1}^{m} \left(a_k;q\right)_n,
\end{equation}
and $\left(e^{\pm i\theta};q\right)_n = \left(e^{i\theta} ,e^{-i\theta};q\right)_n$.

Using $q$-Pochhammer symbols we can define $q$ analogs of many functions, for example the $q$-binomial coefficient is
\begin{equation}
    {n \brack k}_q =  \frac{(q;q)_n}{(q,q)_{n-k} (q;q)_k}.
\end{equation}
This $q$-binomial coefficient becomes the standard binomial coefficient in the limit $q \ra 1^-$, as is standard for $q$ analogs.

Another useful $q$ analog is the basic hypergeometric series, which is the $q$-analogue of the generalized hypergeometric series:
\begin{equation}
    {}_n \phi_{m} \left[\genfrac..{0pt}{0}{a_1,a_2, \ldots , a_n}{b_1,b_2,\ldots, b_m} ; q ,z \right] = \sum_{k=0}^\infty \frac{\left(a_1,a_2, \ldots , a_n,q \right)_k}{\left(b_1,b_2, \ldots , b_m,q \right)_k} (-1)^{k(1+m-n)} q^{(1+m-n)\binom{k}{2}}\frac{z^k}{(q;q)_k}.
    \label{genhypergeomfn}
\end{equation}
The basic hypergeometric series obeys countless identities (see e.g. \cite{Hypergeometric_Book}). In particular we will use
\begin{equation}
            {}_n \phi_{m} \left[\genfrac..{0pt}{0}{1,a_2, \ldots , a_n}{b_1,b_2,\ldots, b_m} ; q ,z \right] = 1  ,
\end{equation}
the $q$-binomial theorem
\begin{equation} \label{eq:qbinomthrm}
    \begin{aligned}
        {}_1 \phi_{0} \left[\genfrac..{0pt}{0}{a}{\text{---}} ; q ,z \right] &= \sum_{k=0}^\infty \frac{(a;q)_k}{(q;q;)_k}z^k = \frac{(az;q)_\infty}{(z;q)_\infty} ,\\
        {}_1 \phi_{0} \left[\genfrac..{0pt}{0}{q^{-n}}{\text{---}} ; q ,z q^n \right] &= \sum_{k=0}^n {n \brack k}_q (-1)^k q^{\binom{k}{2}} z^k = (z;q)_n ,
    \end{aligned}
\end{equation}
and the identity
\begin{equation} \label{eq:hypgeidnt}
            {}_2 \phi_{1} \left[\genfrac..{0pt}{0}{q^{-n},b}{c} ; q ,q \right] = \frac{(c/b;q)_n}{(c;q)_n} b^n .
\end{equation}

Many useful functions and orthogonal polynomials can be defined using the basic hypergeometric series. The most ubiquitous in the text are the Continuous $q$-Hermite polynomials defined by
\begin{equation}
    H_n(x|q) = e^{in\theta} {}_2 \phi_{0} \left[\genfrac..{0pt}{0}{q^{-n},0}{\text{---}} ; q ,q^{n} e^{-2i\theta} \right] = \sum_{k=0}^n {n \brack k}_q e^{i(n-2k)\theta} , \qquad  x = \cos(\theta).
\end{equation}

These polynomials satisfy the recursion relation 
\begin{equation}
    2x H_n(x|q) = H_{n+1}(x|q) + (1-q^n)H_{n-1}(x|q), \qquad \qquad  H_{0}(x|q) = 1, ~~H_{-1}(x|q) = 0,    
\end{equation}
They also satisfy the orthogonality relations and generating function
\begin{align}
\sum_{k=0}^\infty \frac{H_n(x|q) t^n}{(q;q)_n} &= \frac{1}{\left( t e^{\pm i \theta};q\right)_\infty} ,\\
\sum_{k=0}^\infty \frac{H_n(x|q) H_n(x'|q) t^n}{(q;q)_n} = \frac{\left(t^2;q\right)_\infty }{\left( t e^{\pm i \theta \pm i \theta'};q\right)_\infty} &= \frac{\left(t^2;q\right)_\infty }{\left( t e^{ i \theta +i \theta'},t e^{i \theta -i \theta'},t e^{- i \theta +i \theta'},t e^{-i \theta -i \theta'};q\right)_\infty} ,\\
\int_0^\pi \frac{d\theta}{2\pi} \left( q,e^{\pm 2 i \theta};q\right)_\infty H_n(x|q)H_m(x|q) &= (q;q)_n \delta_{n,m} .
\end{align}

We can also expand products of Hermite polynomials as \cite{szabłowski2013qhermitepolynomialsrelationshipfamilies}
\begin{equation} \label{eq:HHtoH}
    H_{k}(x|q)H_{l}(x|q) = \sum_{j=0}^{\min (k,l)} {k \brack j}_q {l \brack j}_q  (q;q)_j H_{k+l - 2j}(x|q),
\end{equation}

Aside from Hermite polynomials, we will also encounter the Al-Salam-Chihara polynomials
\begin{equation}
    Q_n(x|a,b;q) = \frac{(ab;q)_n}{a^n } {}_3\phi_{2}\left[\genfrac..{0pt}{0}{q^{-n},a e^{\pm i \theta}}{ab,0} ; q ,q \right]
\end{equation}
and the continuous Big $q$-Hermite polynomials
\begin{equation}
    H_n(x|a;q) = e^{in\theta} {}_2 \phi_{0} \left[\genfrac..{0pt}{0}{q^{-n},a e^{i\theta}}{\text{---}} ; q ,q^{n} e^{-2i\theta} \right] = \sum_{k=0}^n {n \brack k}_q \left(a e^{i\theta};q\right)_k e^{i(n-2k)\theta} .
\end{equation}
These polynomials satisfy their own generating functions, recursion relations, and orthogonality, which can be found in the literature \cite{HypergeometricSeries,Hypergeometric_Book,szabłowski2013qhermitepolynomialsrelationshipfamilies}. These polynomials are also limits of each other:
\begin{equation}
    H_n(x|a;q) = Q_n(x|a,0;q), \qquad \qquad 
    H_n (x|q) =  H_n(x|0;q) = Q_n(x|0,0;q).
\end{equation}

We can expand the Big $q$-Hermite polynomial as a sum of $q$-Hermite polynomials \cite{szabłowski2013qhermitepolynomialsrelationshipfamilies}
\begin{equation} \label{eq:bigHtoH}
    H_n(x|a;q) = \sum_{k=0}^n {n \brack k}_q (-a)^k q^{\binom{k}{2}} H_{n-k}(x|q).
\end{equation}
Additionally, we will use the property that
\begin{equation} \label{eq:bigHxx}
    H_{n}(x|e^{-i\theta};q) = e^{in\theta} {}_2 \phi_{0} \left[\genfrac..{0pt}{0}{q^{-n},1}{\text{---}} ; q ,q^{n} e^{-2i\theta} \right] = e^{in\theta},
\end{equation}
and similarly $H_{n}(x|e^{i\theta};q) =e^{-i n\theta}$.

We will also use the relation \cite{szabłowski2013qhermitepolynomialsrelationshipfamilies,Roberts_2018}
\begin{equation} \label{eq:sumHHtoQ}
\begin{aligned}
    \sum_{k \geq 0} \frac{t^k H_{k+m}(x_1|q) H_{k+n}(x_2|q)}{\left(q;q\right)_{k}}
     &= \frac{\left(t^2;q \right)_{\infty}}{\left(te^{\pm i \theta_1 \pm i \theta_2 };q \right)_{\infty}} Q_{m,n}(x_1,x_2|t,q),\\
     Q_{m,n}(x_1,x_2|t,q)=& \sum_{s = 0}^n (-1)^s q^{s(s-1)/2} {n \brack s}_q \frac{t^s}{\left(t^2;q\right)_{m+s}} H_{n-s}(x_2|q)Q_{m+s}(x_1|te^{\pm i\theta_2};q).
\end{aligned}
\end{equation}

Finally, we can take the $\theta$ integral in some cases explicitly, writing certain objects as a Neumann series. Of particular interest is the integral \cite{Micha2018}
\begin{equation} \label{eq:intHtoI}
    \int_0^\pi \frac{d\theta}{2\pi} \left(q,e^{\pm 2 i \theta};q \right)_{\infty} e^{- \beta E(\theta)} H_k(\theta)
    = \frac{\sqrt{1-q}}{\beta}\sum_{j=0}^\infty (-1)^j\frac{(q;q)_{k+j}}{(q;q)_j} q^{\frac{j(j+1)}{2}} (k + 2j+1) I_{k+2j+1}\left(\frac{2 \beta}{\sqrt{1-q}} \right)
\end{equation}

We can derive this relation using the expansion \cite{szabłowski2013qhermitepolynomialsrelationshipfamilies,Micha2018}
\begin{equation}
\begin{aligned}
    x^n &= \frac{1}{2^n}\sum_{m=0}^{\lfloor \frac{n}{2}\rfloor} c_{m,n} H_{n-2m}(x|q),\\
    c_{m,n} &= \sum_{j=0}^m (-1)^j q^{j + \binom{j}{2}} \frac{n-2m+2j+1}{n+1}\binom{n+1}{m-j} {n -2m +j \brack j}_q.
\end{aligned}
\end{equation}

Then we can derive \eqref{eq:intHtoI} as
\begin{equation}
    \begin{aligned}
        \int_0^\pi\frac{d\theta}{2\pi} &\left(q,e^{\pm 2 i \theta};q \right)_{\infty} e^{- \beta E(\theta)} H_k(\cos\theta|q)= \\
        &= \sum_{n \geq 0} \frac{1}{n!}\left( \frac{-2\beta}{\sqrt{1-q}}\right)^n\int_0^\pi \frac{d\theta}{2\pi} \left(q,e^{\pm 2 i \theta};q \right)_{\infty} \cos^n\theta H_k(\cos \theta|q) \\
        &= \sum_{n \geq 0} \frac{1}{n!}\left( \frac{-\beta}{\sqrt{1-q}}\right)^n\int_0^\pi \frac{d\theta}{2\pi} \left(q,e^{\pm 2 i \theta};q \right)_{\infty} \sum_{m = 0}^{\lfloor n/2 \rfloor} c_{n,m} H_{n-2m}(\cos \theta|q) H_k(\cos \theta|q) \\
        &= \sum_{n \geq 0} \frac{1}{n!}\left( \frac{-\beta}{\sqrt{1-q}}\right)^n \sum_{m = 0}^{\lfloor n/2 \rfloor} c_{n,m} (q;q)_k \delta_{k,n-2m} \\
        &= (q;q)_k \sum_{n \geq 0} \sum_{m = 0}^{\lfloor n/2 \rfloor} \frac{  \delta_{k,n-2m}}{n!}\left( - \tilde{\beta} \right)^n \sum_{j=0}^{m} (-1)^j q^{j + \binom{j}{2}} \frac{n - 2m +2j +1}{n+1} \binom{n+1}{m-j} {n -2m +j \brack j}_q \\
        &=  \sum_{n \geq 0} \sum_{m = 0}^{\lfloor n/2 \rfloor}  \sum_{j=0}^{m} \frac{  \delta_{k,n-2m}}{n!}\left( - \tilde{\beta} \right)^n (-1)^j q^{j + \binom{j}{2}} \frac{k +2j +1}{n+1} \binom{n+1}{m-j} \frac{(q;q)_{k+j}}{(q;q)_j} \\
        &=   \sum_{m = 0}^{\infty}  \sum_{j=0}^{m} \frac{ 1}{(2m+k)!}\left( - \tilde{\beta} \right)^{2m+k} (-1)^j q^{j + \binom{j}{2}} \frac{k +2j +1}{2m+k+1} \binom{2m+k+1}{m-j} \frac{(q;q)_{k+j}}{(q;q)_j} \\
        &=   \sum_{j=0}^{\infty} (-1)^j q^{j + \binom{j}{2}} \frac{(q;q)_{k+j}}{(q;q)_j}  (k +2j +1) \\
        &\times \sum_{l=m-j = 0}^{\infty}   \frac{ 1}{(2(l+j)+k)!}\left( - \tilde{\beta} \right)^{2(l+j)+k} \frac{1}{2(l+j)+k+1} \binom{2(l+j)+k+1}{l}  \\
        &= \frac{1}{\tilde{\beta}}\sum_{j=0}^{\infty} (-1)^j q^{j + \binom{j}{2}} \frac{(q;q)_{k+j}}{(q;q)_j}  (k +2j +1) I_{k +2j +1}(2\tilde{\beta}) ,
    \end{aligned}
\end{equation}
where $\tilde{\beta} = \frac{\beta}{\sqrt{1-q}}$ and $I_{k}(x)$ is the modified Bessel function of the first kind. 

We can improve slightly on \eqref{eq:intHtoI} by using the recurrence relation of the Bessel functions
\begin{equation}
    \frac{2\nu}{x}I_\nu(x) = I_{\nu-1}(x) - I_{\nu+1}(x) .
\end{equation}
This allows us to write 
\begin{equation} \label{eq:intHtoI2}
    \begin{aligned}
        \int_0^\pi\frac{d\theta}{2\pi} \left(q,e^{\pm 2 i \theta};q \right)_{\infty} e^{- \beta E(\theta)} &H_k(\cos\theta|q) = \frac{1}{\tilde{\beta}}\sum_{j=0}^{\infty} (-1)^j q^{j + \binom{j}{2}} \frac{(q;q)_{k+j}}{(q;q)_j}  (k +2j +1) I_{k +2j +1}(2\tilde{\beta})\\
        &= \sum_{j=0}^{\infty} (-1)^j q^{j + \binom{j}{2}} \frac{(q;q)_{k+j}}{(q;q)_j} \left[ I_{k +2j}(2\tilde{\beta}) - I_{k +2(j+1)}(2\tilde{\beta})\right] \\
        &= \sum_{j=0}^{\infty}  I_{k +2j}(2\tilde{\beta})  (-1)^j q^{\binom{j}{2}} \frac{(q;q)_{k+j-1}}{(q;q)_j} \left[q^j(1-q^{k+j}) + (1-q^j) \right] \\
        &= \sum_{j=0}^{\infty} (-1)^j q^{\binom{j}{2}} \frac{(q;q)_{k+j-1}}{(q;q)_j} \left(1-q^{k+2j}\right) I_{k +2j}(2\tilde{\beta}) .
    \end{aligned}
\end{equation}
This version will be useful in the next appendix.

\section{Expanding \texorpdfstring{$G_\text{late}$}{G late} in Bessel functions} \label{app:glate}

We want to expand the SFF in terms of Bessel functions, to allow us to plot it. Starting from \eqref{eq:Glate}, and using \eqref{eq:intHtoI2} we can write
\begin{equation}
    \begin{aligned}
         G_{late}(\beta_1,\beta_2;q) &=2^{-N} \sum_{k=0}^{\infty} C_k F_{\beta_1,k}(q) F_{\beta_2,k}(q), \\
         F_{\beta,k}(q) &= \sum_{l=0}^\infty \frac{1}{(q;q)_l} \int_0^\pi \frac{d\theta}{2\pi} \left(q,e^{\pm 2 i \theta};q \right)_{\infty} e^{- \beta E(\theta)} H_{k+2l}(\cos \theta |q) \\
         &= \sum_{l=0}^\infty \sum_{j=0}^\infty (-1)^j q^{\binom{j}{2}} \frac{(q;q)_{k+2l+j-1}}{(q;q)_j (q;q)_l}  \left(1-q^{k+2l+2j} \right) I_{k+2l+2j}\left(2\tilde{\beta} \right) \\
         &= \sum_{m=l+j=0}^\infty  \left(1-q^{k+2m} \right) I_{k+2m}\left(2\tilde{\beta} \right) \sum_{j=0}^m (-1)^j q^{\binom{j}{2}} \frac{(q;q)_{k+2m-j-1}}{(q;q)_j (q;q)_{m-j}}.
    \end{aligned}
\end{equation}
Now we can use the properties of the $q$-Pochhammer symbol \eqref{eq:qpochprops}, and the $q$-binomial theorem \eqref{eq:qbinomthrm} to re-write $F_{\beta,k}$ as:
\begin{equation}
    \begin{aligned}
         F_{\beta,k}(q) &= \sum_{m=0}^\infty  \frac{(q;q)_\infty \left(1-q^{k+2m} \right)}{(q;q)_m} I_{k+2m}\left(2\tilde{\beta} \right) \sum_{j=0}^m {m \brack j}_q (-1)^j q^{\binom{j}{2}} \frac{1}{(q^{k+2m-j};q)_\infty}\\
         &= \sum_{m=0}^\infty  \frac{(q;q)_\infty \left(1-q^{k+2m} \right)}{(q;q)_m} I_{k+2m}\left(2\tilde{\beta} \right) \sum_{j=0}^m {m \brack j}_q (-1)^j q^{\binom{j}{2}} \sum_{l=0}^\infty \frac{q^{l(k+2m-j)}}{(q;q)_l}\\
         &= \sum_{m=0}^\infty  \frac{(q;q)_\infty \left(1-q^{k+2m} \right)}{(q;q)_m} I_{k+2m}\left(2\tilde{\beta} \right) \sum_{l=0}^\infty \frac{q^{l(k+2m)}(q^{-l};q)_{m}}{(q;q)_l}.
    \end{aligned}
\end{equation}
Now we notice that $(q^{-l};q)_{m} = 0$ for $l<m$, so we can shift $l \ra l+m$. This results in
\begin{equation}
    \begin{aligned}
         F_{\beta,k}(q) &= \sum_{m=0}^\infty  \frac{(q;q)_\infty \left(1-q^{k+2m} \right)}{(q;q)_m} I_{k+2m}\left(2\tilde{\beta} \right) \sum_{l=0}^\infty \frac{q^{(l+m)(k+2m)}(q^{-m-l};q)_{m}}{(q;q)_{m+l}} \\
         &= \sum_{m=0}^\infty  \frac{(q;q)_\infty \left(1-q^{k+2m} \right)}{(q;q)_m} I_{k+2m}\left(2\tilde{\beta} \right) \sum_{l=0}^\infty \frac{q^{(l+m)(k+2m)}}{(q;q)_{l}} (-1)^m q^{\binom{m}{2} - m(m+l)}  \\ 
         &= \sum_{m=0}^\infty  \frac{(q;q)_\infty}{(q^{m+k};q)_\infty (q;q)_m} (-1)^m q^{\binom{m}{2} + m (k+m)} \left(1-q^{k+2m} \right) I_{k+2m}\left(2\tilde{\beta} \right) \\
         &=\sum_{m=0}^\infty  \left(q^{m+1};q\right)_{k-1} (-1)^m q^{\binom{m}{2} + m (k+m)} \left(1-q^{k+2m} \right) I_{k+2m}\left(2\tilde{\beta} \right) .
    \end{aligned}
\end{equation}

Thus we can summarize the result 
\begin{equation}
    \begin{aligned}
          G_{late}(\beta_1,\beta_2;q) &=2^{-N} \sum_{k=0}^{\infty} C_k F_{\beta_1,k}(q) F_{\beta_2,k}(q), \\
         F_{\beta,k}(q) &=\sum_{m=0}^\infty  \left(q^{m+1};q\right)_{k-1} (-1)^m q^{\binom{m}{2} + m (k+m)} \left(1-q^{k+2m} \right) I_{k+2m}\left(\frac{2\beta}{\sqrt{1-q}} \right) ,
    \end{aligned}
\end{equation}
which is possible to plot.

\section{The coincident limit of \texorpdfstring{$\rho_{\text{late}}$}{} and the late time ramp} \label{sec:Glate}

We wish to take the $\theta_1 \ra \theta_2$ limit of $\rho_{\text{late}}$ given in \eqref{eq:rholate}, which should contain the universal behavior encoding the RMT ramp \cite{Cotler:2016fpe}. To evaluate this limit, it is useful to consider a regularized version of this sum by adding a small weight $t = 1-\epsilon$ to intersections with the marked chord. While this sum does make sense as a distribution without need for regularization (equivalently the integrals in \eqref{eq:Glate} are finite,) we wish to take the coincident limit of the smooth part of $\rho_{\text{late}}$, and thus regularizing the sum in  \eqref{eq:rholate} ensures the result is smooth. The regularized version $\rho_{\text{late}}$ is 

\begin{equation} \label{eq:rholatereg}
\begin{aligned}
    \rho_{late}(\theta_1,\theta_2;t,q) &= 2^{-N} \frac{\left(q,q,e^{\pm 2i\theta_1},e^{\pm 2i\theta_2};q\right)_\infty}{(2\pi)^2} \sum_{r,l_1,l_2=0, s>0}^\infty t^{r+s} \frac{H_{r+s+2l_1}(x_1|q) H_{r+s+2l_2}(x_2|q)}{(q;q)_{r}(q;q)_{s} (q;q)_{l_1} (q;q)_{l_2}}.
\end{aligned}
\end{equation}

Next, we can take the sum over $r$ using \eqref{eq:sumHHtoQ}, resulting in 
\begin{equation} \label{eq:rholatereg2}
\begin{aligned}
    \rho_{late}(\theta_1,\theta_2;t,q) &= 2^{-N} \frac{\left(q,q,e^{\pm 2i\theta_1},e^{\pm 2i\theta_2};q\right)_\infty}{(2\pi)^2} \sum_{l_1,l_2=0, s>0}^\infty t^{s} \frac{(t^2;q)_\infty}{(te^{\pm i\theta_1\pm i \theta_2};q)_{\infty}}\frac{Q_{s+2l_1,s+2l_2}(x_1,x_2|t,q)}{(q;q)_{s} (q;q)_{l_1} (q;q)_{l_2}},
\end{aligned}
\end{equation}
where $Q_{m,n}(x_1,x_2|t,q)$  is defined in \eqref{eq:sumHHtoQ}.

Now we can take the limit $\theta_1 \ra \theta_2$ of \eqref{eq:rholatereg2} near $t = 1$. To take this limit it is useful to use the identities $(t^2,q)_m =(1-t^2)(t^2q,q)_{m-1}$ and $\left(e^{\epsilon};q\right)_\infty \approx (1-e^\epsilon) ( q;q)_\infty$. Then we have that near $t=1$ and $\theta_1 \approx \theta_2$
\begin{equation} \label{eq:tqlim}
\begin{aligned}
     \frac{\left(t^2;q \right)_{\infty}}{\left(te^{\pm i \theta_1 \pm i \theta_2 };q \right)_{\infty}} &Q_{j,l}(x_1,x_2|t,q) =  \frac{\left(t^2;q \right)_{\infty}}{\left(te^{\pm i \theta_1 \pm i \theta_2 };q \right)_{\infty}} \sum_{s = 0}^{l}  (-1)^s q^{s(s-1)/2} {l \brack s}_q t^s \\ &\qquad  \times H_{l-s}(x_2|q) \frac{1}{t^{j+s} e^{i(j+s)\theta_2}} \sum_{m=0}^{j+s} \frac{\left(q^{-j-s},t e^{i\theta_2\pm i \theta_1} ;q\right)_m}{\left(t^2 ;q\right)_{m}} \frac{q^m}{(q;q)_m}\\
     & \approx  \frac{(1-t e^{-i (\theta_1 - \theta_2)})}{4\left(q,e^{\pm i (\theta_1 + \theta_2) };q \right)_{\infty}\sin^2((\theta_1-\theta_2)/2)} \sum_{s = 0}^{l}  (-1)^s q^{s(s-1)/2} {l \brack s}_q  
     \\ \qquad &\qquad \times H_{l-s}(x_2|q) \frac{1}{ e^{i(j+s)\theta_2}} \sum_{m=1}^{j+s} \left(q^{-j-s}, e^{i(\theta_2+ \theta_1)} ;q\right)_m  \frac{q^m}{(q;q)_m} \\
     &\qquad +  \frac{(1-t^2)\left(q;q \right)_{\infty}}{\left(e^{\pm i \theta_1 \pm i \theta_2 };q \right)_{\infty}} \sum_{s = 0}^{l}  (-1)^s q^{s(s-1)/2} {l \brack s}_q H_{l-s}(x_2|q) e^{-i(j+s)\theta_2},
     \end{aligned}
\end{equation}
where in the last line we separated the $m=0$ case from the rest. Next we can split
\begin{equation}
    \frac{\left(t^2;q \right)_{\infty}}{\left(te^{\pm i \theta_1 \pm i \theta_2 };q \right)_{\infty}} Q_{j,l}(x_1,x_2|t,q) \approx A + B,
\end{equation}
where $A$ is the first two lines in the result of \eqref{eq:tqlim} while $B$ is last line in \eqref{eq:tqlim}. Then we can use \eqref{eq:hypgeidnt},\eqref{eq:bigHtoH} and \eqref{eq:bigHxx} to simplify this limit
\begin{equation}
\begin{aligned}
     A &= \frac{(1-t e^{-i (\theta_1 - \theta_2)})}{4\left(q,e^{\pm i (\theta_1 + \theta_2) };q \right)_{\infty}\sin^2((\theta_1-\theta_2)/2)} \sum_{s = 0}^{l}  (-1)^s q^{s(s-1)/2} {l \brack s}_q  \\ \qquad & \qquad\times H_{l-s}(x_2|q) \frac{1}{ e^{i(j+s)\theta_2}} \left[{}_2\phi_1 \left[\genfrac..{0pt}{0}{q^{-j-s},e^{i(\theta_2+ \theta_1)}}{0} ; q ,q \right] -1 \right]\\
     &= \frac{(1-t e^{-i (\theta_1 - \theta_2)})}{4\left(q,e^{\pm i (\theta_1 + \theta_2) };q \right)_{\infty}\sin^2((\theta_1-\theta_2)/2)} \sum_{s = 0}^{l}  (-1)^s q^{s(s-1)/2} {l \brack s}_q  \\ \qquad & \qquad\times H_{l-s}(x_2|q)  \left(e^{i(j+s)\theta_1} - e^{-i(j+s)\theta_2}\right) \\
     & =  \frac{(1-t e^{-i (\theta_1 - \theta_2)})}{4\left(q,e^{\pm i (\theta_1 + \theta_2) };q \right)_{\infty}\sin^2((\theta_1-\theta_2)/2)}\\
     &\qquad \times\left(e^{ij \theta_1} H_{l}\left(x_2|e^{i\theta_1};q \right) - e^{- ij \theta_2} H_{l}\left(x_2|e^{-i\theta_2};q \right) \right)\\
     &\approx  \frac{(1-t)}{4\left(q,e^{\pm i (\theta_1 + \theta_2) };q \right)_{\infty}\sin^2((\theta_1-\theta_2)/2)} \left(e^{i (j-l) (\theta_1 +\theta_2)/2} - e^{-i (j-l) (\theta_1 +\theta_2)/2} \right),
\end{aligned}    
\end{equation}
and 
\begin{equation}
    \begin{aligned}
        B &=  \frac{(1-t^2)\left(q;q \right)_{\infty}}{\left(e^{\pm i \theta_1 \pm i \theta_2 };q \right)_{\infty}} \sum_{s = 0}^{l}  (-1)^s q^{s(s-1)/2} {l \brack s}_q H_{l-s}(x_2|q) e^{-i(j+s)\theta_2} \\
        & = \frac{(1-t^2)\left(q;q \right)_{\infty}}{\left(e^{\pm i \theta_1 \pm i \theta_2 };q \right)_{\infty}} e^{i(l-j)\theta_2}\\
        &\approx \frac{(1-t)}{2\left(q,e^{\pm i (\theta_1 + \theta_2) };q \right)_{\infty}\sin^2\left((\theta_1-\theta_2)/2\right)} e^{i(l-j)(\theta_1+\theta_2)/2} .
    \end{aligned}
\end{equation}
Note that the approximation $\theta_1 \approx \theta_2$ allows us to write $2\theta_1 \approx 2\theta_2 \approx \theta_1+\theta_2$.

Calling $\Delta \theta \equiv \theta_1-\theta_2 \ll 1$, and $\bar{\theta} \equiv (\theta_1 + \theta_2)/2 \approx \theta_1 \approx \theta_2$, we have using the above approximations
\begin{equation}
    \begin{aligned}
        \rho_\text{late}(\bar{\theta},\Delta\theta;q) & \approx \lim_{t \ra 1}- 2^{-N} (1-t) \frac{\left(q,e^{\pm 2i\bar{\theta}};q\right)_\infty}{4(2\pi)^2 \sin^2(\Delta \theta /2)} \\
        & \qquad \qquad \times \sum_{l_1,l_2=0}^\infty \sum_{s = 1}^\infty t^s \frac{e^{i2(l_1-l_2)\bar{\theta}} - e^{-i2(l_1-l_2)\bar{\theta}} + 2 e^{-i2(l_1-l_2)\bar{\theta}}}{\left( q,q\right)_{l_1} \left( q,q\right)_{l_2} \left( q,q\right)_s}\\
        &= \lim_{t \ra 1} - 2^{-N}(1-t) \frac{\left(q,e^{\pm 2i\bar{\theta}};q\right)_\infty}{8\pi^2 \sin^2(\Delta \theta /2)} \frac{1}{\left(e^{\pm 2i \bar{\theta}};q\right)_\infty} \left( \frac{1}{(t;q)_{\infty}}-1\right)\\
        &= - \frac{2^{-N} }{8\pi^2 \sin^2(\Delta \theta /2)},
    \end{aligned}
\end{equation}
where we used the $q$-binomial theorem, \eqref{eq:qbinomthrm}, to evaluate the sums over $l_1,l_2,$ and $s$.

From this approximation we can compute the late time ramp as
\begin{equation} \label{eq:glateramp}
    \begin{aligned}
        G_{\text{late}}(\beta,t \gg 1, q) &= \int d\theta_1 d\theta_2 \rho_\text{late}(\theta_1,\theta_2) e^{-\beta(E(\theta_1)+E(\theta_2)) +it(E(\theta_1)-E(\theta_2))}\\
        &\approx - 2^{-N} \frac{1}{8\pi^2} \int  \frac{d\theta_1 d\theta_2}{\sin^2((\theta_1-\theta_2) /2)} e^{-\frac{4 \beta}{\sqrt{1-q}}\cos((\theta_1+\theta_2)/2) +i\frac{4 t}{\sqrt{1-q}}\sin((\theta_1+\theta_2)/2)\sin((\theta_1-\theta_2)/2)}\\
        &\approx - 2^{-N} \frac{1}{4\pi^2} \int  \frac{d\theta_+ d\theta_-}{\theta_-^2} e^{-\frac{4 \beta}{\sqrt{1-q}}\cos(\theta_+) +i\frac{4 t}{\sqrt{1-q}}\theta_-\sin(\theta_+)}\\
        &= 2^{-N} \frac{1}{4\pi^2} \int_0^\pi  d\theta_+ \frac{4 \pi t \sin(\theta_+)}{\sqrt{1-q}} e^{-\frac{4 \beta}{\sqrt{1-q}}\cos(\theta_+)}\\
        &= 2^{-N} \frac{1}{2 \pi \beta} \sinh \left(\frac{4 \beta}{\sqrt{1-q}}\right)~ t.
    \end{aligned}
\end{equation}

Note that the final integral in \eqref{eq:glateramp} has the form
\begin{equation}
    G_{\text{late}}(\beta,t \gg 1, q) =2^{-N} \frac{t}{2\pi} \int_{-E_0}^{E_0} dE ~e^{-2 \beta E} .
\end{equation}
This is precisely the form of the universal ramp (see e.g. equation (42) in \cite{Cotler:2016fpe}.)

\clearpage
\bibliographystyle{JHEP}
\bibliography{DSSYK}

\end{document}